\documentclass[twocolumn]{aastex63}

\received{XXX}
\revised{XXX}
\accepted{XXX}
\submitjournal{AAS Journals}

\shorttitle{MAPS VIII: Gap Chemistry in AS 209}
\shortauthors{Alarc\'on et al.}
\graphicspath{{./}{figures/}}

\begin{document}

\title{Molecules with ALMA at Planet-forming Scales (MAPS) VIII: CO Gap in AS 209--Gas Depletion or Chemical Processing?}

\correspondingauthor{Felipe Alarc\'on}
\email{falarcon@umich.edu}

\author[0000-0002-2692-7862]{Felipe Alarc\'on }
\affiliation{Department of Astronomy, University of Michigan,
323 West Hall, 1085 S University, Ave.,
Ann Arbor, MI 48109, USA}

\author[0000-0003-4001-3589]{Arthur D. Bosman}
\affiliation{Department of Astronomy, University of Michigan,
323 West Hall, 1085 S University, Ave.,
Ann Arbor, MI 48109, USA}

\author[0000-0003-4179-6394]{Edwin A. Bergin}
\affiliation{Department of Astronomy, University of Michigan,
323 West Hall, 1085 S University, Ave.,
Ann Arbor, MI 48109, USA}

\author[0000-0002-0661-7517]{Ke Zhang}
\altaffiliation{NASA Hubble Fellow}
\affiliation{Department of Astronomy, University of Wisconsin-Madison, 
475 N Charter St, Madison, WI 53706}
\affiliation{Department of Astronomy, University of Michigan, 
323 West Hall, 1085 S University, Ave., 
Ann Arbor, MI 48109, USA}

\author[0000-0003-1534-5186]{Richard Teague}
\affiliation{Center for Astrophysics \textbar\, Harvard \& Smithsonian, 60 Garden St., Cambridge, MA 02138, USA}

\author[0000-0001-7258-770X]{Jaehan Bae}
\altaffiliation{NASA Hubble Fellowship Program Sagan Fellow}
\affil{Earth and Planets Laboratory, Carnegie Institution for Science, 5241 Broad Branch Road NW, Washington, DC 20015, USA}
\affiliation{Department of Astronomy, University of Florida, Gainesville, FL 32611, USA}

\author[0000-0003-3283-6884]{Yuri Aikawa}
\affiliation{Department of Astronomy, Graduate School of Science, The University of Tokyo, Tokyo 113-0033, Japan}

\author[0000-0003-2253-2270]{Sean M. Andrews} 
\affiliation{Center for Astrophysics \textbar\, Harvard \& Smithsonian, 60 Garden St., Cambridge, MA 02138, USA}

\author[0000-0003-2014-2121]{Alice S. Booth} \affiliation{Leiden Observatory, Leiden University, 2300 RA Leiden, the Netherlands}
\affiliation{School of Physics and Astronomy, University of Leeds, Leeds, LS2 9JT, UK}

\author[0000-0002-0150-0125]{Jenny K. Calahan} 
\affiliation{Department of Astronomy, University of Michigan,
323 West Hall, 1085 S University, Ave.,
Ann Arbor, MI 48109, USA}

\author[0000-0002-2700-9676]{Gianni Cataldi}
\affiliation{National Astronomical Observatory of Japan, 2-21-1 Osawa, Mitaka, Tokyo 181-8588, Japan}
\affil{Department of Astronomy, Graduate School of Science, The University of Tokyo, Tokyo 113-0033, Japan}

\author[0000-0002-1483-8811]{Ian Czekala}
\altaffiliation{NASA Hubble Fellowship Program Sagan Fellow}
\affiliation{Department of Astronomy and Astrophysics, 525 Davey Laboratory, The Pennsylvania State University, University Park, PA 16802, USA}
\affiliation{Center for Exoplanets and Habitable Worlds, 525 Davey Laboratory, The Pennsylvania State University, University Park, PA 16802, USA}
\affiliation{Center for Astrostatistics, 525 Davey Laboratory, The Pennsylvania State University, University Park, PA 16802, USA}
\affiliation{Institute for Computational \& Data Sciences, The Pennsylvania State University, University Park, PA 16802, USA}
\affiliation{Department of Astronomy, 501 Campbell Hall, University of California, Berkeley, CA 94720-3411, USA}

\author[0000-0001-6947-6072]{Jane Huang}
\altaffiliation{NASA Hubble Fellowship Program Sagan Fellow}
\affiliation{Department of Astronomy, University of Michigan, 323 West Hall, 1085 S University, Ave., Ann Arbor, MI 48109, USA}
\affiliation{Center for Astrophysics \textbar\, Harvard \& Smithsonian, 60 Garden St., Cambridge, MA 02138, USA}

\author[0000-0003-1008-1142]{John D. Ilee}
\affiliation{School of Physics and Astronomy, University of Leeds, Leeds, LS2 9JT, UK}

\author[0000-0003-1413-1776]{Charles J. Law}
\affiliation{Center for Astrophysics \textbar\, Harvard \& Smithsonian, 60 Garden St., Cambridge, MA 02138, USA}

\author[0000-0003-1837-3772]{Romane Le Gal}
\affiliation{Center for Astrophysics \textbar\, Harvard \& Smithsonian, 60 Garden St., Cambridge, MA 02138, USA}
\affiliation{IRAP, Universit\'{e} de Toulouse, CNRS, CNES, UT3, 31400 Toulouse, France}
\affiliation{Univ. Grenoble Alpes, CNRS, IPAG, F-38000 Grenoble, France}
\affiliation{IRAM, 300 rue de la piscine, F-38406 Saint-Martin d'H\`{e}res, France}

\author[0000-0002-7616-666X]{Yao Liu}\affiliation{Purple Mountain Observatory \& Key Laboratory for Radio Astronomy, Chinese Academy of Sciences, Nanjing 210023, China}

\author[0000-0002-7607-719X]{Feng Long}
\affiliation{Center for Astrophysics \textbar\, Harvard \& Smithsonian, 60 Garden St., Cambridge, MA 02138, USA}

\author[0000-0002-8932-1219]{Ryan A. Loomis}\affiliation{National Radio Astronomy Observatory, 520 Edgemont Rd., Charlottesville, VA 22903, USA}

\author[0000-0002-1637-7393]{Fran\c cois M\'enard}
\affiliation{Univ. Grenoble Alpes, CNRS, IPAG, F-38000 Grenoble, France}

\author[0000-0001-8798-1347]{Karin I. \"Oberg} \affiliation{Center for Astrophysics \textbar\, Harvard \& Smithsonian, 60 Garden St., Cambridge, MA 02138, USA}

\author[0000-0002-6429-9457]{Kamber R. Schwarz} \altaffiliation{NASA Hubble Fellowship Program Sagan Fellow}
\affiliation{Lunar and Planetary Laboratory, University of Arizona, 1629 E University Blvd, Tucson, AZ 85721, USA}

\author[0000-0002-2555-9869]{Merel L. R. van 't Hoff}
\affiliation{Department of Astronomy, University of Michigan,
323 West Hall, 1085 S University, Ave.,
Ann Arbor, MI 48109, USA}

\author[0000-0001-6078-786X]{Catherine Walsh}
\affiliation{School of Physics and Astronomy, University of Leeds, Leeds, LS2 9JT, UK}

\author[0000-0003-1526-7587]{David J. Wilner}
\affiliation{Center for Astrophysics \textbar\, Harvard \& Smithsonian, 60 Garden St., Cambridge, MA 02138, USA}




\begin{abstract}
Emission substructures in gas and dust are common in protoplanetary disks. Such substructures can be linked to planet formation or planets themselves. We explore the observed gas substructures in AS 209 using thermochemical modeling with \texttt{RAC2D} and high-spatial resolution data from the Molecules with ALMA at Planet-forming Scales(MAPS) program. The observations of C$^{18}$O $J=2-1$ emission exhibit a strong depression at 88 au overlapping with the positions of multiple gaps in millimeter dust continuum emission.
We find that the observed CO column density is consistent with either gas surface-density perturbations or chemical processing, while C$_2$H column density traces changes in the C/O ratio rather than the H$_2$ gas surface density. However, the presence of a massive planet ($>$ 0.2 M$_{Jup}$) would be required to account for this level of gas depression, which conflicts with constraints set by the dust emission and the pressure profile measured by gas kinematics. Based on our models, we infer that a local decrease of CO abundance is required to explain the observed structure in CO, dominating over a possible gap-carving planet present and its effect on the H$_2$ surface density. This paper is part of the MAPS special issue of the Astrophysical Journal Supplement Series.

\end{abstract}



\section{Introduction} \label{sec:intro}

Surveys of protoplanetary disks have shown that substructures in the dust continuum emission are ubiquitous \citep[e.g.][]{DSHARPI,Long..et..al..2018,Cieza..et..al..2019}. The presence of substructures such as gaps and rings is seen as a signpost of incipient and ongoing planet formation that alters the local disk's physical conditions \citep{Pinilla..et..al..2012,Dong..et..al..2015}. A natural question is whether the emission of common observable species traces the local disks chemistry or physical changes caused by planet--disk interactions.

The AS 209 system is an ideal laboratory to test the difference in gas emission substructure caused by planet--disk interactions and local chemistry as it presents multiple substructures in dust continuum and line emission \citep{Jane..2016, DSHARPII,Vivi..et..al2018,law20_rad}. These substructures, in particular the millimeter dust continuum emission, have been previously analyzed for a possible association with hidden planets \citep{Dong..et..al..2018,Fedele..et..al..2018,Favre..et..al..2019}.  We analyze the gas chemical and physical state using high-resolution ($\sim$0$\farcs$15) ALMA observations of C$^{18}$O and C$_2$H from Molecules with ALMA at Planet-forming Scales \citep{czekala20,oberg20} \footnote{\href{www.alma-maps.info}{www.alma-maps.info}}. The C$^{18}$O emission profile shows a wide emission depression centered at 88 au with a width of 47 au, while C$_2$H shows an emission ring centered at 70 au with a width of 68 au \citep{law20_rad}. The gap in C$^{18}$O emission translates into a decrease in the CO column density, being depleted by at least 47\% when compared with a smooth surface--density profile \citep{zhang20_maps}. Because of its relatively simple chemistry and strong lines \citep{Kamp..et..al..2011,Williams..Best..2015,Molyarova..et..al..2017}, CO is frequently used as a H$_2$ gas tracer assuming a uniform abundance across the disk. Therefore, such CO depletion can be explained by the presence of a sub-Jovian planet carving a H$_2$ gap \citep{Favre..et..al..2019}.

In general, local CO abundance variations at substructures are considered to be a minor problem compared to the question of the global CO abundance relative to H$_2$ and its possible deviations from the interstellar medium(ISM) value (e.g. $\sim$10$^{-4}$), necessary for the use of CO emission as a calibrated tracer of the disk gas mass \citep[][]{Ted..Williams..2018}. However, detailed models suggest that standard scaling between CO and H$_2$ surface density does not hold locally inside the dust substructures due to CO chemical processing \citep{Nienke..et..al..2018,Alarcon..et..al..2020}.



Given the limitations of CO emission as a tool in probing disk structure in planet-forming regions, additional chemical tracers observed in the MAPS ALMA Large Program  provide new constraints between planet--disk interactions and local disk chemistry cases \citep{oberg20}. In this regard, the emission from C$_2$H stands out as its emission is often as bright as $^{13}$CO in some disk systems \citep[][]{Kastner14} and C$_2$H emission is widely detected in gas-rich systems often exhibiting emission rings \citep{Kastner15, Ted..et..al..2016,Bergner..et..al..2019}. Numerical models suggest that the explanation for such high C$_2$H column densities is a localized high C/O ratio \citep{Cleeves..et..al..2018,Miotello..et..al..2019} and a photon-dominated chemistry necessary for C$_2$H production \citep{Ted..et..al..2016}.

In this work, we analyze two main scenarios explaining the CO column density structure, CO chemical processing or H$_2$ gas depletion caused by a giant planet. Our aim is to disentangle the degeneracy between CO abundance and H$_2$ surface density via a combined observational and numerical study of AS 209, understanding the possible footprint of a newborn planet. Thus, we explore a solution where chemical processing locally alters the CO abundance across the dust gaps and another scenario in which planet--disk interactions carve a gap in the H$_2$ surface density assuming a constant CO abundance. We also explore whether the conditions producing structure in the CO surface density also lead to concurrent production of C$_2$H.

This paper is organized as follows: We describe the setup of our thermochemical models in Section \ref{sec:methods}. Section \ref{sec:CO} presents the degeneracy between CO abundance and gas surface density, and Section \ref{sec:C2H} shows the role of C$_2$H as a tracer of active chemistry. Then, we discuss the main conclusions of our work in Section \ref{sec: Discussion}. A brief summary of our work is presented in Section \ref{sec: Summary}.



\section{Chemical and Physical Modeling} \label{sec:methods}

In our goal of reproducing the CO and C$_2$H column-density profiles in AS 209, we use a chemical and a physical approach to understand the origin of the line-emission substructure. Our  chemistry-dominated approach assumes that a depleted CO abundance is required to reproduce structure in C$^{18}$O emission in the context of a smooth H$_2$ gas surface-density profile.  The physical approach assumes a constant CO abundance and that the structure is the result of a local decrease in the H$_2$ surface density. We built the models so they match the inferred CO column densities, and we compared the expected column density of C$_2$H and other tracers of the chemical and physical structure of the disk, such as emission heights and kinematic deviation, to disentangle the degeneracy between both solutions to the CO radial profile.

Additionally, we introduced a variable C/O ratio to both models keeping the CO abundance constant as well. We also account for multiple dust substructures in the disk for both large and small dust grains.  For completeness, we provide a detailed review of the AS 209 system as a planet formation laboratory in Appendix \ref{sec: AS209}.

\subsection{Chemical Network and Thermochemical Code}

We used the 2D thermochemical code \texttt{RAC2D}, described in \cite{Fujun..Ted..RAC2D}, to model the  chemical evolution in AS 209 for 1 Myr and to reproduce the inferred CO and C$_2$H column densities in the disk. The chemical network includes 524 species with 6425 chemical reactions. It uses the reaction rates from the UMIST 2012 database \citep{McElroy..et..al..2012} for the gas-phase chemistry. \texttt{RAC2D} also includes reactions that take place on the surface of dust grains using the formalisms of \cite{Hasegawa..1992}. For the photodesorption of H$_2$O and OH by Ly$\alpha$ photons, the code uses yield values from \cite{Oberg..et.al..2009} and the yield values from \cite{Oberg..2009b} for CO$_2$ and CO.

\subsection{AS 209 Model}

The AS 209 disk structure includes multiple gaps and rings beyond 20 au. We adopted the disk parameters for AS 209 from \cite{zhang20_maps} with small changes due to the inclusion of an inner gap at 7 au with a 90\% depletion of dust and H$_2$ gas, and a width of 3.8 au. We list the parameter of our models in Table \ref{tab:diskpars}. The location and depletion of this inner gap is uncertain. However, the analysis of \cite{bosman20_inner20au} using the MAPS data of AS 209 requires a strong H$_2$ gas depletion in the inner 10 au of at least one order of magnitude, which is consistent with the presence of a gap or a gas cavity. 

\begin{deluxetable}{l|l|l|l}
\tablecaption{Parameters for the AS 209 disk model.  \label{tab:diskpars}}
\tablehead{
\colhead{Parameter} & \colhead{Value} &\colhead{Parameter} & \colhead{Value} 
}
\startdata
Disk mass & 0.0045 $M_{\odot}$ & $r_c$  & 80 au  \\
Large dust mass & 4 x 10$^{-4}$ $M_{\odot}$ &  $\gamma$  & 1.0\\
Small dust mass &  5 x 10$^{-5}$ $M_{\odot}$ & $h_c$ & 6 au  \\ $\zeta$ ionization rate  & 1.36 x  10$^{-18}$ s$^{-1}$
& $\psi$ & 1.25 \\
\enddata
\end{deluxetable}

The grid in our models is defined on the $(r,z)$ plane. We have a radial logarithmic spacing with 250 radial cells ranging from 1 to 250 au covering our radial range of interest, which is between 20 to 120 au. The vertical spacing in the grid is variable and depends on the local physical conditions of the disk, with a spatial refinement  in the layers where CO and C$_2$H have abundances with respect to H  higher than 10$^{-7}$ and 10$^{-13}$, respectively. The cells are never larger than 3 au.

\subsubsection{Gas Structure}

\begin{figure}
    \centering
    \includegraphics[width=0.5\textwidth]{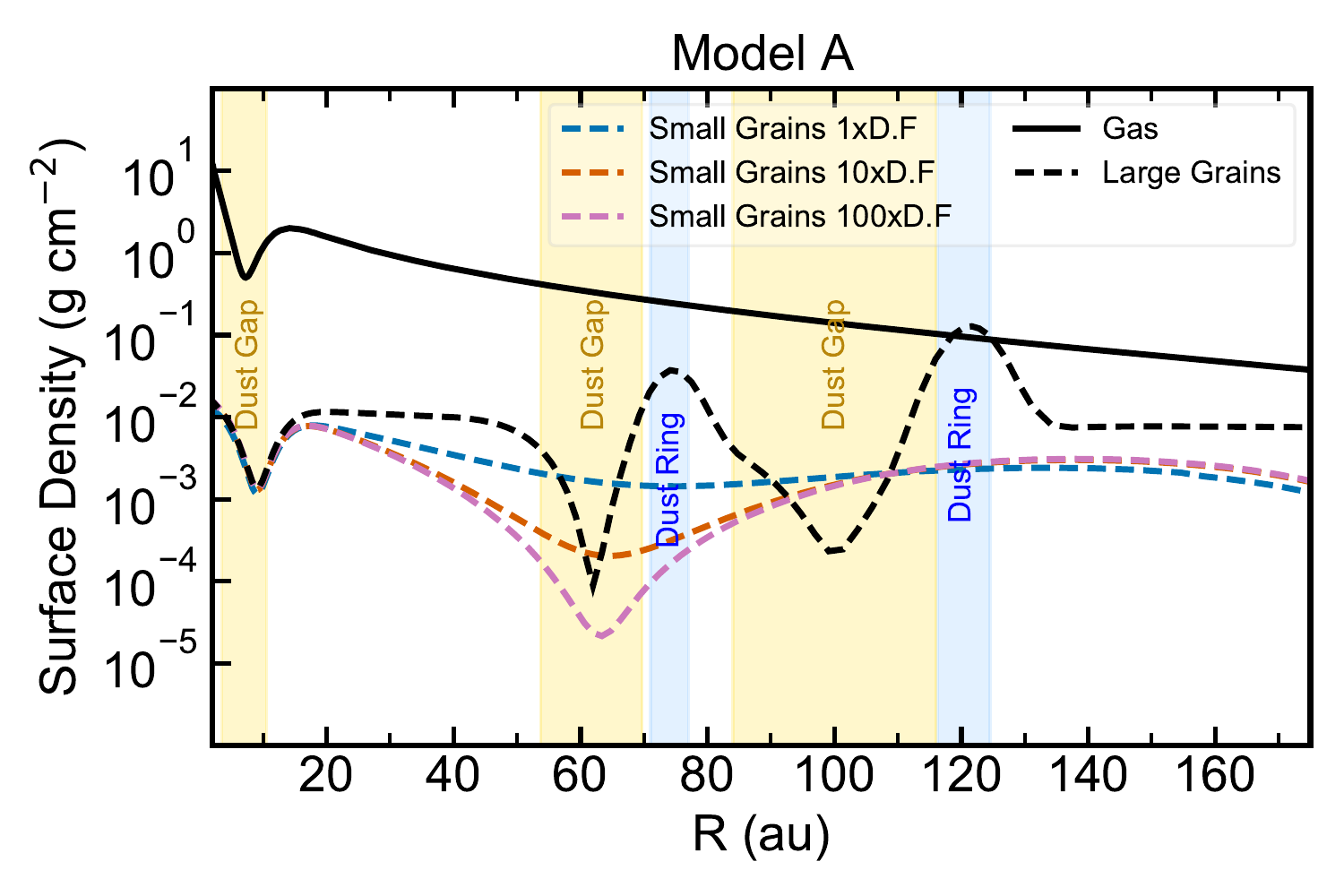}
    \includegraphics[width=0.5\textwidth]{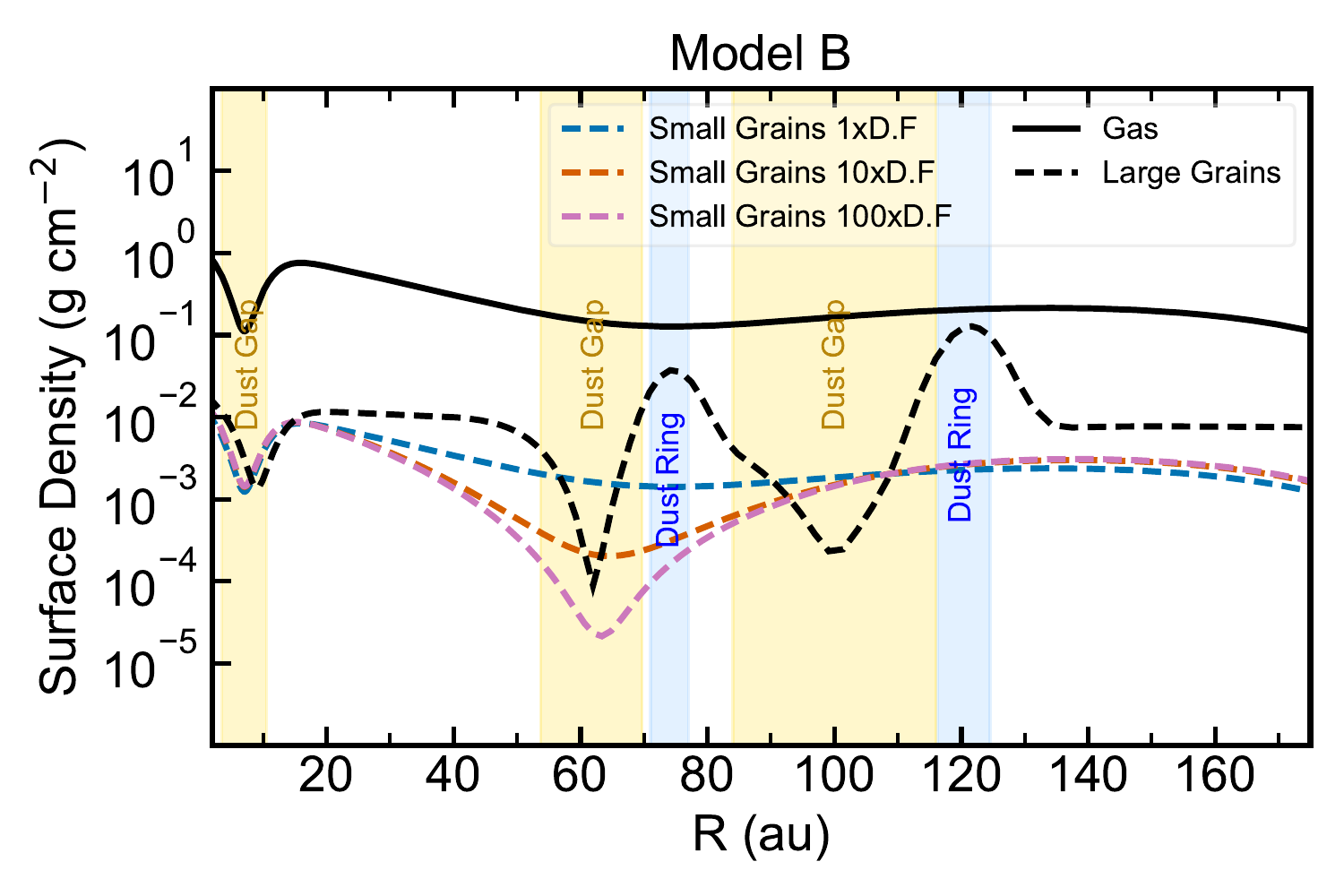}
    \caption{Two sets of models with varying surface density for the small grains. The small-grain surface density was changed by varying the depletion in the 59 au gap with three different depletion factors, 1x (nominal), 10x and 100x. \textbf{Top}: In Model A, the surface density has no gas substructures, i.e., it follows a smooth H$_2$ surface-density profile, where CO is not a H$_2$ mass tracer. \textbf{Bottom}: Model B, showing the surface density using the CO structure as a H$_2$ gas mass tracer, scaling with a constant CO abundance. Model B shows a wide gas gap at 59 au. The shaded areas in yellow and blue are the dust gaps and dust ring, respectively.}
    \label{fig:Sigma_gas}
\end{figure}

We modeled the disk density structure using two different approaches, calling them Model A and Model B.  For Model A we used a smooth self-similar distribution \citep{Lynden-Bell_1974} for the gas surface-density profile, i.e,


\begin{equation}\label{eq: Self-Similar}
    \Sigma_{\rm smooth}(r) = \Sigma_c \left(\frac{r}{r_c} \right)^{-\gamma}\exp \left[-\left(\frac{r}{r_c} \right)^{2-\gamma}\right],
\end{equation}

\noindent where $r_c$  is the characteristic radius for the disk and $\Sigma_c$ is the characteristic surface-density of the disk. Figure \ref{fig:Sigma_gas} shows the gas density structure in Models A and B  as well as the dust surface-density structure, which is explained further in Section \ref{Sect: Dust}. The main difference between the surface-density profiles in Models A and B is the presence of a H$_2$ surface-density drop scaled from the inferred CO column density at the location of the C$^{18}$O emission gap. We show the thermal structure and the differences between Models in Appendix \ref{app:Model}.

We set the vertical distribution with a Gaussian centered at the midplane \citep{armitage_2020}:

\begin{equation}\label{eq: vertical Dist}
\rho(r,z) = \rho_0(r)\exp\Big(-\frac{z^2}{2h^2} \Big),
\end{equation}

\noindent with $\rho_0 = \frac{\Sigma(r)}{\sqrt{2\pi}h}$ and $h$ being the scale height, which is different for the gas and the dust. The parameters and gas distribution of Model A are the same as the one used in \cite{zhang20_maps}. The scale height also has a radial dependence through a power-law with a flaring index $\psi$:

\begin{equation}\label{eq: Flaring}
h(r) = h_{c}\Big(\frac{r}{r_{c}} \Big)^{\psi},
\end{equation}
\noindent with $\psi$ the flaring index and $h_c$ the characteristic scale height at the characteristic radius, $r_c$. 


\begin{figure}
    \centering
    \includegraphics[width=0.5\textwidth]{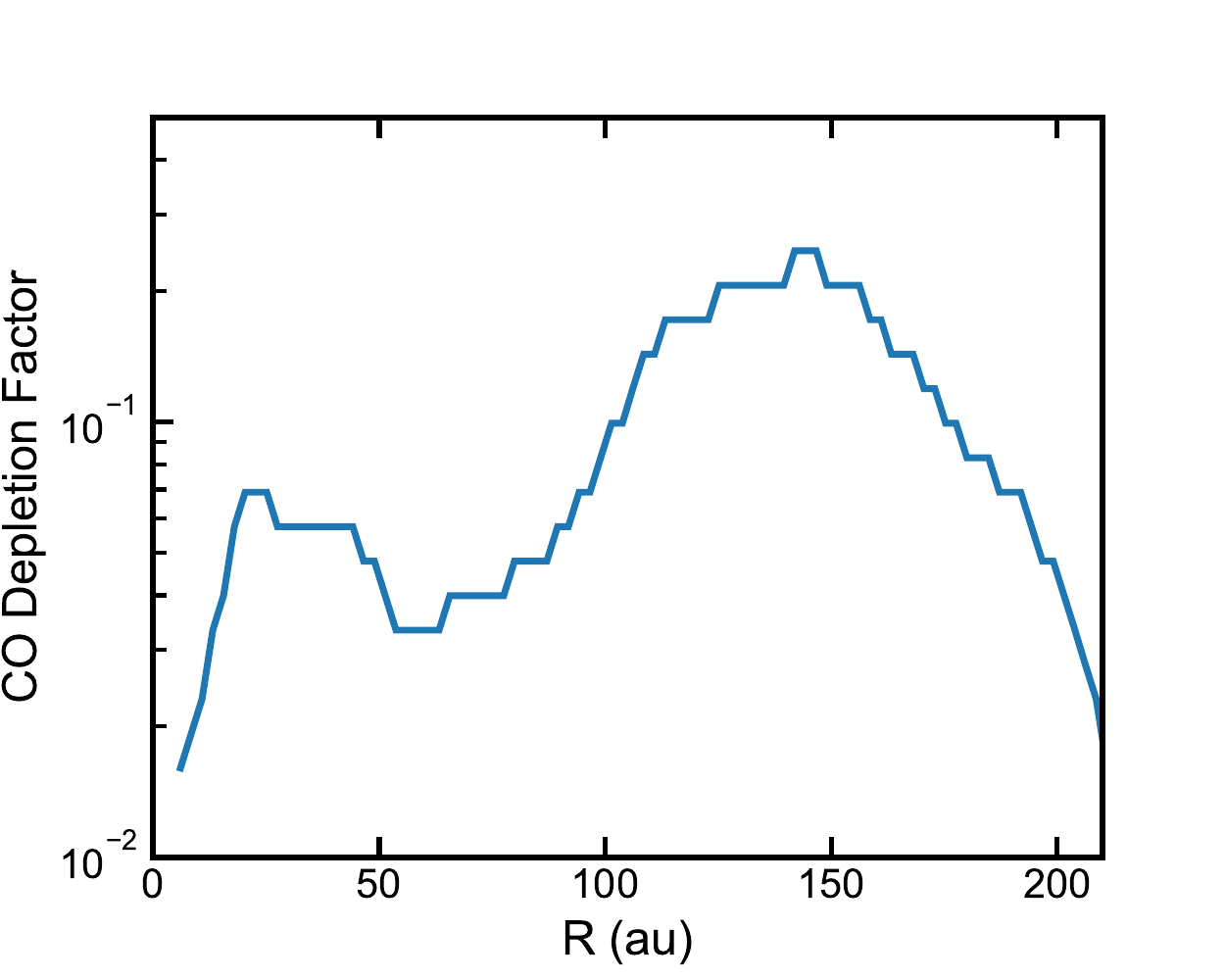}
    \caption{CO depletion used in Model A to fit the CO intensity profile. This depletion is used to compare the case with CO chemical processing (Model A) and the case with CO used as a tracer of H$_2$ surface density (Model B).}
    \label{fig:CO_depletion}
\end{figure}

In Model A we assume a smooth surface-density profile for the H$_2$ gas.
To account for the C$^{18}$O $J=2-1$ emission depression at $\sim$80 au, we require a localized change in the CO abundance.  Here, we adopt the CO depletion profile in Figure \ref{fig:CO_depletion}, i.e., the CO abundance is locally weighted according to that profile. This depletion profile implies that the CO chemical processing in the disk is taken into account effectively as an initial condition. 

In Model B, we scaled the observed CO column-density profile to a H$_2$ column-density profile. We assumed a constant CO abundance of $2.8\times 10^{-4}$  and then calculated the disk chemical evolution with that gas mass distribution. In this scenario, the structure is caused principally by dynamical evolution of the disk rather than chemically evolution of the CO reservoir. Model B implicitly assumes that the H$_2$ gas surface density has a wide gap ($w_{\rm gap}$= 13 au) at 59 au. As a benchmark, we also compare whether or not Model B matches the observed column densities for C$_2$H and the effects the gas depletion may have on the C$_2$H chemical equilibrium.

The assumed initial abundances for Model B with respect to the total number of hydrogen atoms are provided in Table \ref{tab:Abundances}. In order to match the CO column density and the constraints given by the disk mass, our simulations started with a global depletion of C and O that is reduced by one order of magnitude compared to ISM values \citep{Nieva..Przybilla..2012}. We did not include any atomic carbon or water ice initially as they will be added to change the C/O ratio locally for further exploration of the C$_2$H chemistry.

\begin{deluxetable}{l|l|l|l}
\tablecaption{Initial Abundances of the simulations with respect to the number of H Atoms. \label{tab:Abundances}}

\tablehead{
\colhead{Species} & \colhead{Abundance} &\colhead{Species} & \colhead{Abundance} 
}
\startdata
H$_2$ & 5(-1) & He  & 9(-2)  \\
HD  & 2(-5) &  S & 8(-9) \\
CO & 2.8(-4) & N & 7.5(-5)\\
Si$^+$ & 8(-9) & Na$^+$ & 2(-9)  \\
Mg$^+$  & 7(-9) & P & 3(-9) \\
Fe$^+$ & 3(-9) & F & 2(-9)  \\
Cl & 4(-9) & & \\
\enddata
\tablecomments{All the abundances are written in the standard form A(B) = A$\times 10^{B}$}
\tablenotemark*{In our simulations we have C/O=1. We only added atomic carbon or water ice to change the C/O ratio locally, either to increase it or decrease it respectively.}
\end{deluxetable}

\subsubsection{Dust Structure}\label{Sect: Dust}

We have two dust populations in our models. The opacity of each dust population is set by its composition and respective sizes. The large dust population consists of dust grains with sizes $a$ ranging from 0.005 $\mu$m to 1 mm with a dust size distribution following the standard size distribution from \cite{Mathis..et..al..1977} and a total mass of 4.5x10$^{-4}$ $M_{\odot}$.  The large grain population is composed of a mixture of 40\% refractory organics \citep{Henning..1996}, 33\% silicates \citep{Draine..2003}, 20\% water \citep{Warren..and..Brandt} and 7\% troilite \citep{Henning..1996} by mass. The X-ray opacities follow the prescription from \cite{Bethell..Bergin}, which depends on the size (cross section) of each dust grain. The surface-density distribution for the large dust is different from that for the gas and the small dust. In the large dust-surface-density  distribution, the substructures, i.e., gaps and rings, were added into the underlying dust surface-density profile with the self-similar solution profile \citep{Lynden-Bell_1974}. 
Both large dust rings have Gaussian shapes, as in \cite{Alarcon..et..al..2020}, and their amplitudes were chosen so that the mass in the rings matches the inferred mass from \cite{Dullemond..DSHARP}; see Table \ref{tab:pars} for the substructure parameters in our models. The large dust surface density has a lower $\gamma$, 0.2 instead of 1. This $\gamma$ is different from our gas and small-dust distribution as it provides a better fit for the dust continuum emission. A lower $\gamma$  in the large dust grains has a steeper cutoff, implying  that the pebble or millimeter disk is more concentrated in the inner regions as they suffer from more efficient radial drift \citep{Birnstiel..et..al..2010,Birnstiel..et..al..2012}. We used a different approach for the pebble disk than \cite{zhang20_maps}. We used a prescription that analytically characterizes the gaps and rings using previous models from \cite{Fedele..et..al..2018} and \cite{Dullemond..DSHARP}, instead of the iterative approach of \cite{zhang20_maps}.

\begin{deluxetable*}{l|l|l|l}
\tablecaption{Parameters for the substructures present in the dust surface density of Millimeter-sized particles in our models. \label{tab:pars}} 
\tablehead{
\colhead{Substructure} & \colhead{Location (au)} &\colhead{Width (au)} & \colhead{Depletion / Enhancement} 
}
\startdata
Gap 1 & 7 & 3.8  & 0.9 \\
Gap 2  & 62  & 8  & 0.991 \\
Ring 1 & 74 & 3.5 & 7.3  \\
Gap 3 & 100 & 16 & 0.975\\
Ring 2  & 120 & 4.11 & 26 \\
\enddata
\tablenote*{We use values from DSHARP \citep{DSHARPI,Dullemond..DSHARP}. The prescription used for the substructure is the same as in \cite{Alarcon..et..al..2020}, where the width and depletion/enhancement represent the width of a gaussian and the amplitude of the modulation of the surface density.}
\end{deluxetable*}

The small grains have sizes ranging from 0.005 $\mu$m to 1 $\mu$m with a total mass of 5 x 10$^{-5}$ $M_{\odot}$. We assume that the composition of the small grains is different from that of the large grains with a composition of 50\% silicates and 50\%  refractory organics. Instead of following the large dust distribution, the small grains follow the H$_2$ gas surface density as they are more dynamically coupled with the molecular hydrogen gas. Since the small-dust surface density is less constrained in terms of its mass distribution (in comparison to the large millimeter-sized dust), we used different localized small-dust depletion factors where the CO column-density gap is located to assess their effect on different CO and C$_2$H, in particular their vertical distribution and column density.

For both models, A and B, we ran three different simulations changing the small-dust surface density. The only parameter that changes between each simulation is the depth of the gap at $r=59$ au, while the rest of the features of the gap, i.e. width and radius, remain the same. Changes in the radiation field are expected as the small dust is the main source of UV opacity in the upper layers, playing a significant role in the chemistry of photochemical tracers such as C$_2$H or HCN \citep{Fogel..et..al..2011}. Moreover, small dust also changes the temperature of the disk in intermediate layers as the large dust is more settled and concentrated in the midplane.

\subsection{Variable C/O Ratio}

\begin{figure}
    \centering
    \includegraphics[width=0.5\textwidth]{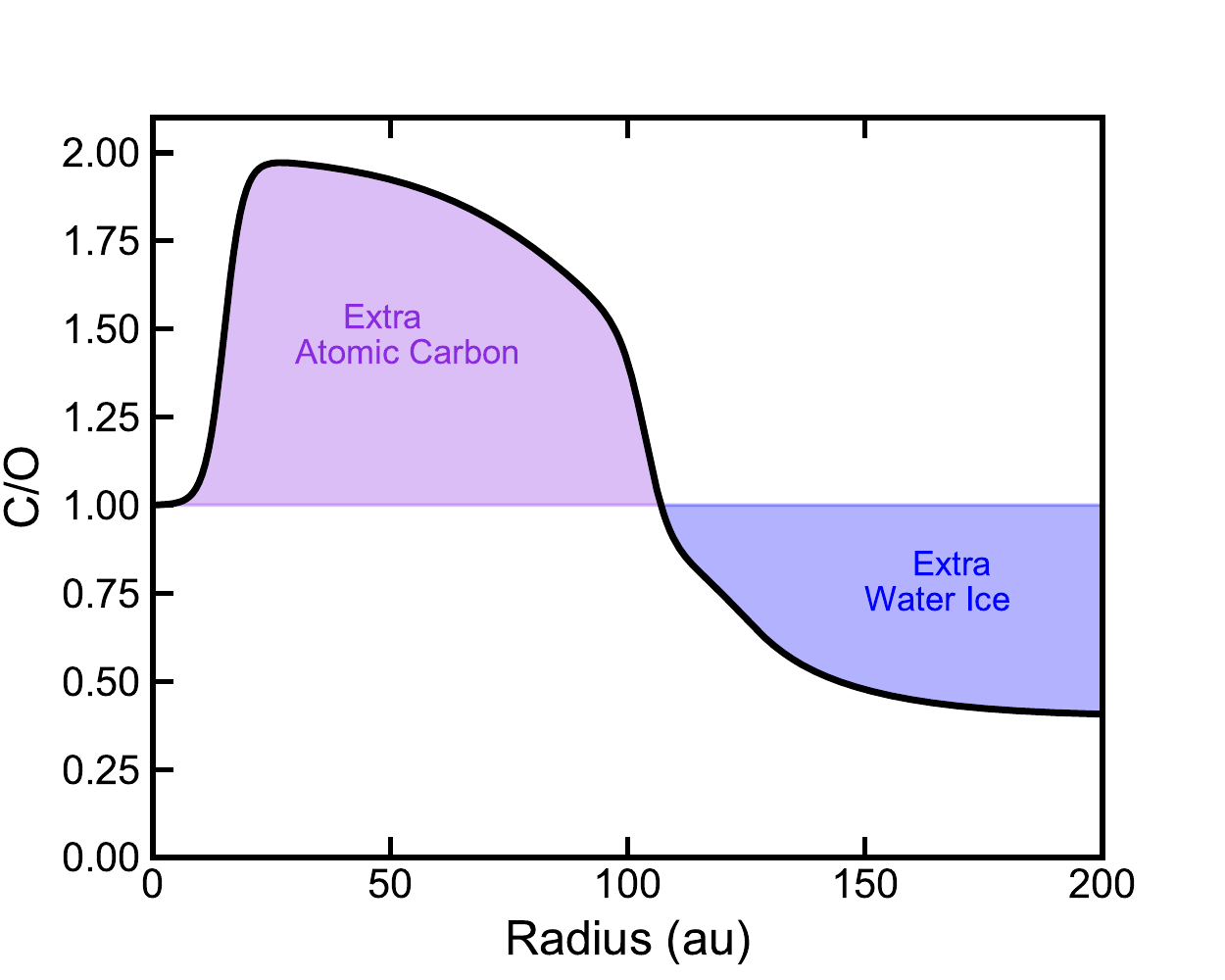}
    \caption{C/O ratio for our models with radial changes in the amount of C. The C/O starts at 1, then it reaches a value of 2 in the gas gap where extra atomic carbon is added. In this region most of the carbon grain destruction would occur. It also coincides with the region where the C$_2$H column density reaches its highest values. Beyond 100 au we added extra water ice to decrease the C/O ratio toward an ISM value C/O=0.4. }
    \label{fig:C_to_Oratio}
\end{figure}

In order to reproduce the high column densities of C$_2$H \citep{bosman20_CtoO,guzman20}, we require deep UV penetration and an increase in the C/O elemental ratio \citep{Ted..et..al..2016, Cleeves..et..al..2018,Miotello..et..al..2019}. For each model, we tried two different approaches for the C/O ratios to find if, under depleted conditions, high column densities at the level observed by \citet{guzman20} are recoverable with just a low C/O ratio. 
In our first approach, we started with C/O=1 instead of 0.4, meaning that the C/O ratio was already higher than the ISM value \citep{Wilson..and..Rood}; this is the expected value for the gas outside the CO ice line, which is found between 15 and 20 au in our model. In the second approach, we added extra carbon to the abundances listed in Table \ref{tab:Abundances}, which are constant in the vertical direction.  We therefore kept the CO abundance constant and added extra neutral atomic carbon in the CO emission gap to increase the C/O ratio to 2. Beyond 100 au, where the C$_2$H emission drops significantly, we added extra water ice  to decrease the C/O ratio to 0.4, as water ice is the main carrier of oxygen in disks \citep{Ewine..2021}. Figure \ref{fig:C_to_Oratio} illustrates the C/O ratio for the models and its radial dependence. The transitions in the radial profile of the C/O ratio were empirically created to fit the C$_2$H inferred column density as best as possible. Irrespective of the location of the C/O ratio transitions, a high C/O ratio is required to match the peak in the C$_2$H column-density profile. In addition to changing elemental abundances, we also reduced the abundance of small grains; this locally increases the UV field promoting the C$_2$H production \citep{Bosman..2021}. 
We then compared our results for the two sets of models, with a particular focus on C$_2$H since CO column density should not have a significant increase as the starting point is C/O=1. 

\subsection{Emission Heights}

We compared our models to the derived column-density structure for a given species.
As an additional constraint, we also compare the emission heights provided in \cite{law20_surf} with the emission heights of CO isotopologue lines from our models. Given the vertical structure in the CO abundance and gas temperature, we used the region where the optical depth $\tau$ of each line ranged between 0.67 and 2 as a proxy of each emitting layer. To produce the different optical depth layers for each molecular transition we used the temperature structure and CO distribution from \texttt{RAC2D} as input for the radiative transfer code \texttt{RADMC-3D} \citep{RADMC3D}.

We generated the emitting heights in the $J=2-1$ rotational lines of the two most abundant CO isotopologues: CO and  $^{13}$CO. As we did not include isotopologue-selective photodissociation, we assumed a constant isotopologue ratio for $^{13}$CO, setting CO $^{13}$CO=69 \citep{Wilson..et..al..1999}. The possible effects of isotopologue-selective photodissociation are further discussed in Section \ref{sec:iso}. The C$^{18}$O $J=2-1$ line is mostly optically thin in AS 209, so the emission height is uncertain. Thus, we did not include it in the analysis of AS 209 in this work.

\section{CO Abundance and H$_2$ Surface-Density Degeneracy} \label{sec:CO}

\subsection{CO column densities}

\begin{figure}
    \centering
    \includegraphics[width=0.5\textwidth]{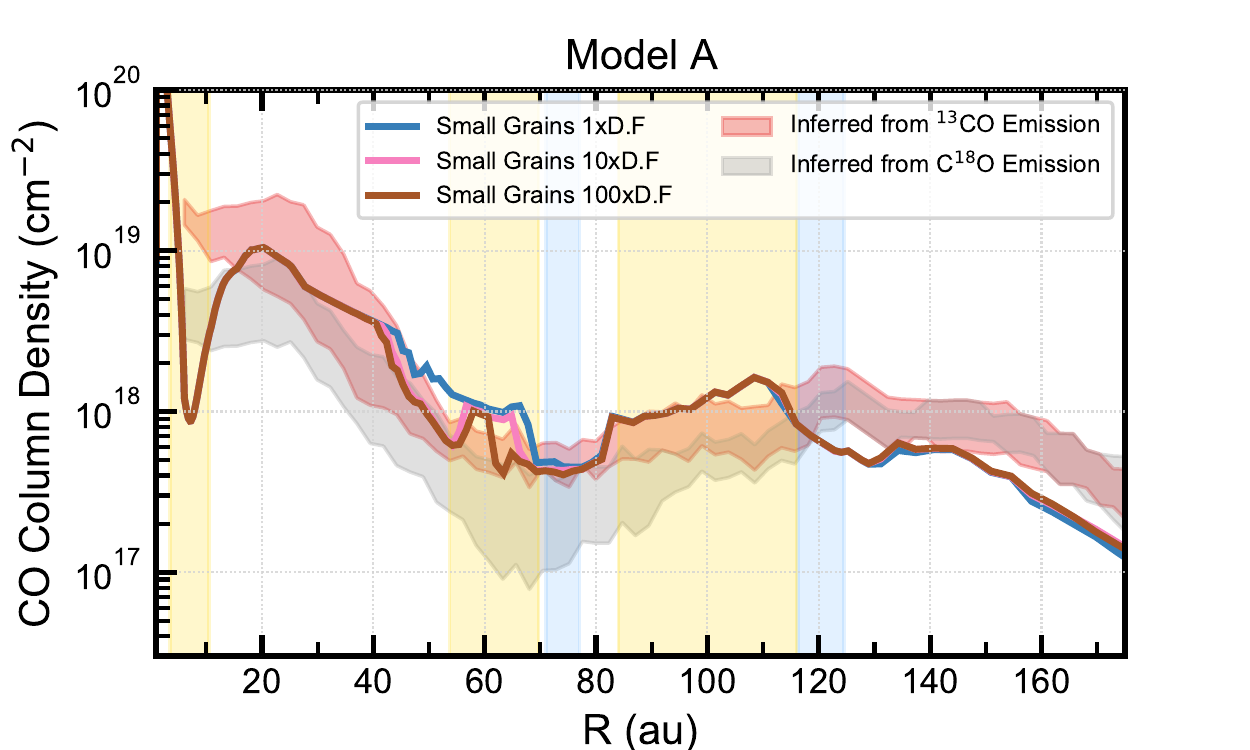}
    \includegraphics[width=0.5\textwidth]{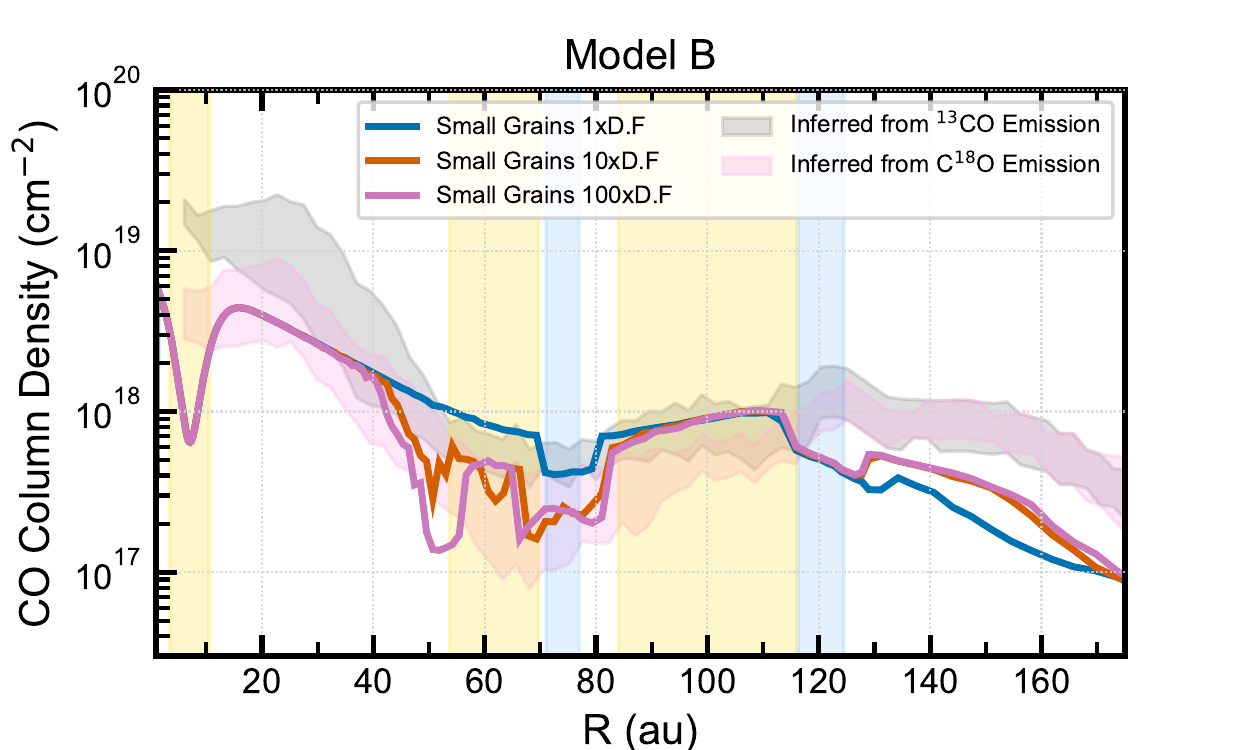}
    \caption{CO column-density profiles for both sets of simulations. \textbf{Top:} smooth gas surface density applying the CO depletion profile in Figure \ref{fig:CO_depletion}. \textbf{Bottom:} Simulation set with CO as a gas mass tracer. The vertical shaded areas are the location of dust substructures, while the radial shaded regions represent the CO column densities derived in \cite{zhang20_maps} using $^{13}$CO and C$^{18}$O emission. In both cases, Models A and B, we observe that whether including CO chemical processing or the gas depletion match the recovered CO column densities. The shaded regions are the locations of the dust gaps and dust rings, in yellow and blue, respectively.}
    \label{fig:Model_CO}
\end{figure}

We show the CO column densities from the thermochemical simulation in Figure \ref{fig:Model_CO}. Both sets of models, A and B, are able to recover the CO column density from \cite{zhang20_maps} with subtle differences, but still within the uncertainties. Therefore, CO column densities alone cannot discriminate between CO processing or the presence of a gas gap carved by a massive planet. As we do not have a full understanding of the CO depletion in the disk, the fact that both models match the CO column-density profiles illustrates the degeneracy between the two solutions to explain the structure in CO emission \citep{Calahan..et..al}.  This is not unexpected, but we confirm that the gas thermal physics does not change substantially between these solutions to favor one model over the other as temperature is more dependent on the dust structure in the midplane and intermediate layers of the disk.  We show the differences between the temperature structures of both in Appendix \ref{app:2d T}; the differences are usually less than 10\%, so they may present some small radial or vertical variation for the abundance of given species, but it does not change the results of our models significantly.  Therefore, we need to look at other features in the emission to break the degeneracy. We provide a deeper discussion of the 2D structure of the models and possible implication for line emission in Appendix~\ref{app:Model}.

By changing the amount of small dust in the gap at 59 au, we do not observe noticeable differences in the CO column densities. However, changes in the CO emission are still expected as the small dust is a key player in setting the disk temperature, particularly in the warm molecular layer on top of the large dust.

\subsection{CO Emission Heights: Model versus Data}

\begin{figure}
    \centering
    \includegraphics[width=0.49\textwidth]{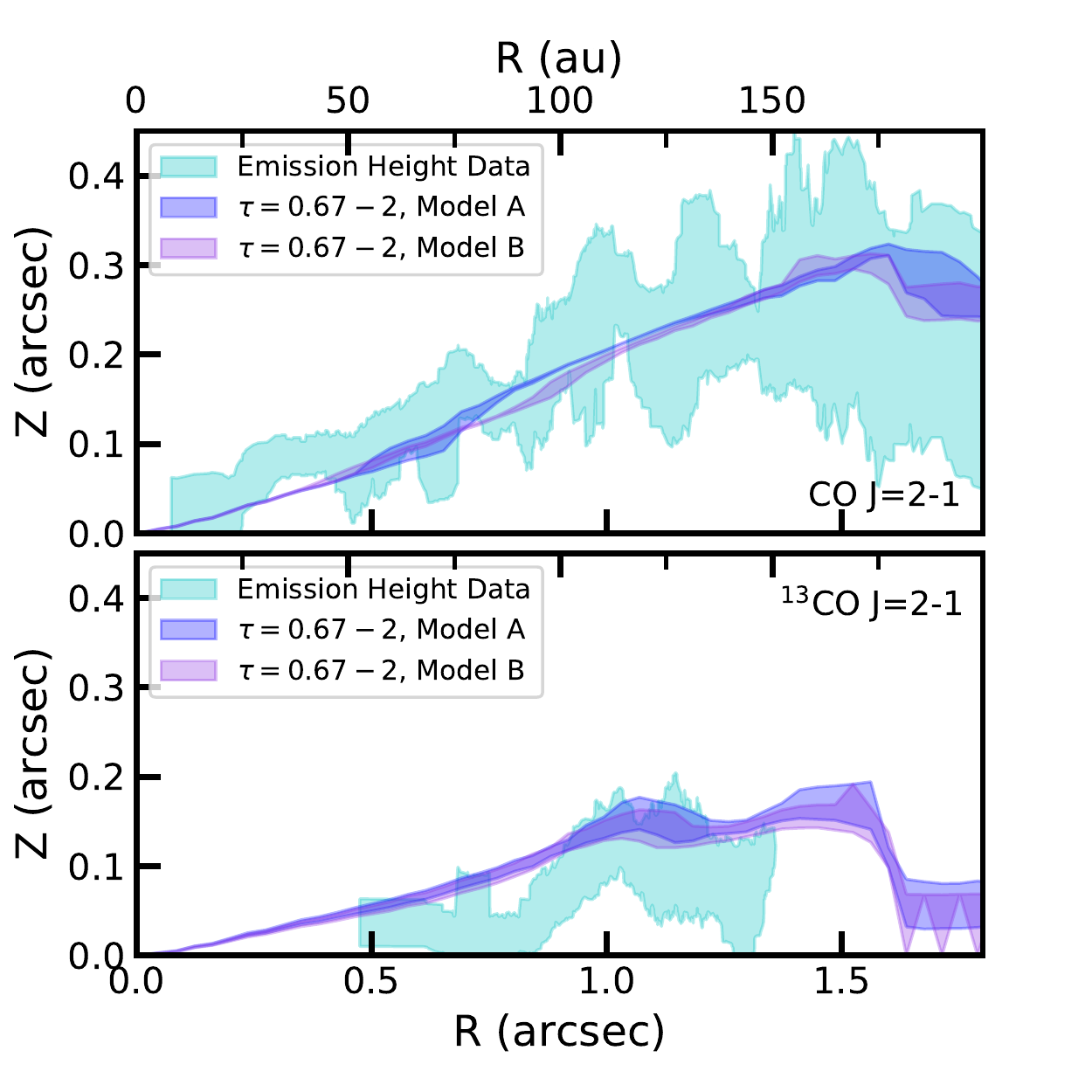}
    \caption{Emitting heights measured in \cite{law20_surf} compared to the layers in the disk where the optical depth ranges from 0.67 to 2 for each line. This layer serves as a probe of the emitting height as most of the contribution of the line should come from these heights. The comparison seems to coincide well with the surfaces of CO and $^{13}$CO. Models A and B are within the range given by the observations. }
    \label{fig:H_emit}
\end{figure}

 The observed emission heights present an additional constraint, in particular for the CO abundance structure.
The comparison between the emitting heights in our models and the ones inferred from the data are shown in Figure \ref{fig:H_emit} \citep{law20_surf}. Here, it is shown that for the optically thick CO, the emitting heights for both models, A and B, match, within the uncertainties, the data. 


For $^{13}$CO, the observed emitting height is constrained in a narrower radial range and closer to the midplane. Nevertheless, Models A and B still agree with the constraints given by the observations. Because both models are within the uncertainties and match emission heights, the degeneracy between these models remains.

\subsection{Pressure Profile and Planets, Breaking the Degeneracy}

CO column densities and emission heights do not distinguish between chemical processing and planet--disk interactions. Therefore, we explore the thermal H$_2$ gas thermal pressure in our models along the CO-emitting  layers and compare them to that obtained by \cite{Teague..et..al..2018} from gas kinematics in AS 209. We expect the pressure profiles to be different around the CO emission gap as the H$_2$ gas surface density is different in Models A and B. 

\subsubsection{Pressure Profiles and Kinematic Deviations}
We made two radial cuts enclosing the emission heights of $^{12}$CO and measured the pressure profile along those layers. One cut is assumed to follow the emitting height:

\begin{equation}\label{eq:surface_Rich}
    h=23\Big(\frac{r}{100} \Big)^{1.05} \mathrm{au},
\end{equation}

\noindent consistent with the layer explored by \cite{Teague..et..al..2018}. The second cut was

\begin{equation}\label{eq:surface_maps}
    h=26.5\left(\frac{r}{121} \right)^{1.292}\exp \left(-\left(\frac{r}{216}\right)^{4.9}\right) \  \mathrm{au},
\end{equation}

\noindent which is the parametric CO $J=2-1$ emission height found by \citep{law20_surf}. A comparison between each layer is illustrated in Figure \ref{fig:HEIGHTS}.

\begin{figure}
    \centering
    \includegraphics[width=0.5\textwidth]{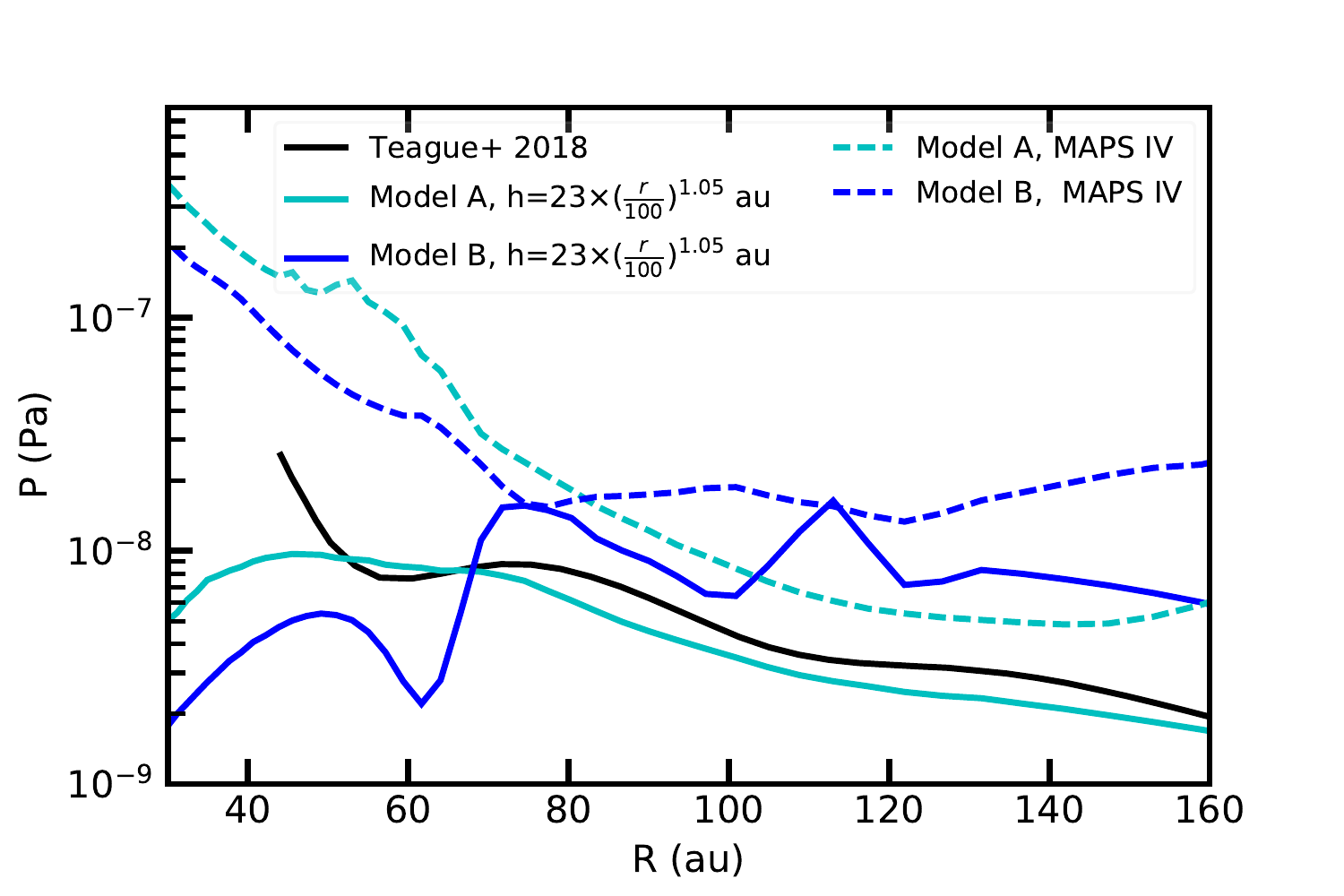}
    \caption{Thermal pressure profile cuts in our two models compared to the profile from \cite{Teague..et..al..2018} inferred from AS 209 disk kinematics using CO line emission. The other cut was taken following the parametric emission surface of CO inferred by \cite{law20_surf}. When the gas is depleted by roughly an order of magnitude, as in Model B, the pressure profile shows significant variations with the radius, these are inconsistent with those from \cite{Teague..et..al..2018}. The lack of strong fluctuations in the observed pressure profile disfavors the presence of a deep gas gap, at least at $\sim$ 59 au, breaking the degeneracy between chemical processing and gas depletion. Thus, chemical processing is the dominant mechanism for CO column variations in AS 209.}
    \label{fig:Pressure_profile}
\end{figure}

We show a comparison between the pressure profiles in our models along the layers given by Eq. \ref{eq:surface_Rich}  and \ref{eq:surface_maps}, and the observed pressure profile from \cite{Teague..et..al..2018} in Figure \ref{fig:Pressure_profile}. As we do not include hydrodynamics in our models, we look at the trend of each model rather than matching the empirically derived pressure profile, which is more directly constrained by its slope and structure than its absolute magnitude. The thermal variation between models, in particular, along the CO emission layer is $\sim$10\% (see Appendix \ref{app:2d T}). Therefore, the variations in the pressure profile are mostly induced by radial density variations, as thermal variations at a 10\% level do not cause significant changes to the pressure profiles. The pressure profile from Model A is smooth. Conversely, Model B exhibits a considerable pressure dip in the region where CO is being depleted in Model A, tracing a one-order-of-magnitude density drop in gas surface density. When we compare our pressure profiles with the one inferred from gas kinematics \citep{Teague..et..al..2018}, the pressure drops seen in Model B are not reproduced.  Instead the pressure analysis suggests a smoother H$_2$ surface-density profile (which could be interpreted as smooth CO changes) as assumed in Model A. We note that the actual pressure values do not exactly match  \citet{Teague..et..al..2018}.  This is particularly the case when we use the emitting surfaces from \cite{law20_surf}. Regardless, radial variation in the pressure profile  probes changes in the H$_2$ gas surface density.

\begin{figure}
    \centering
    \includegraphics[width=0.45\textwidth]{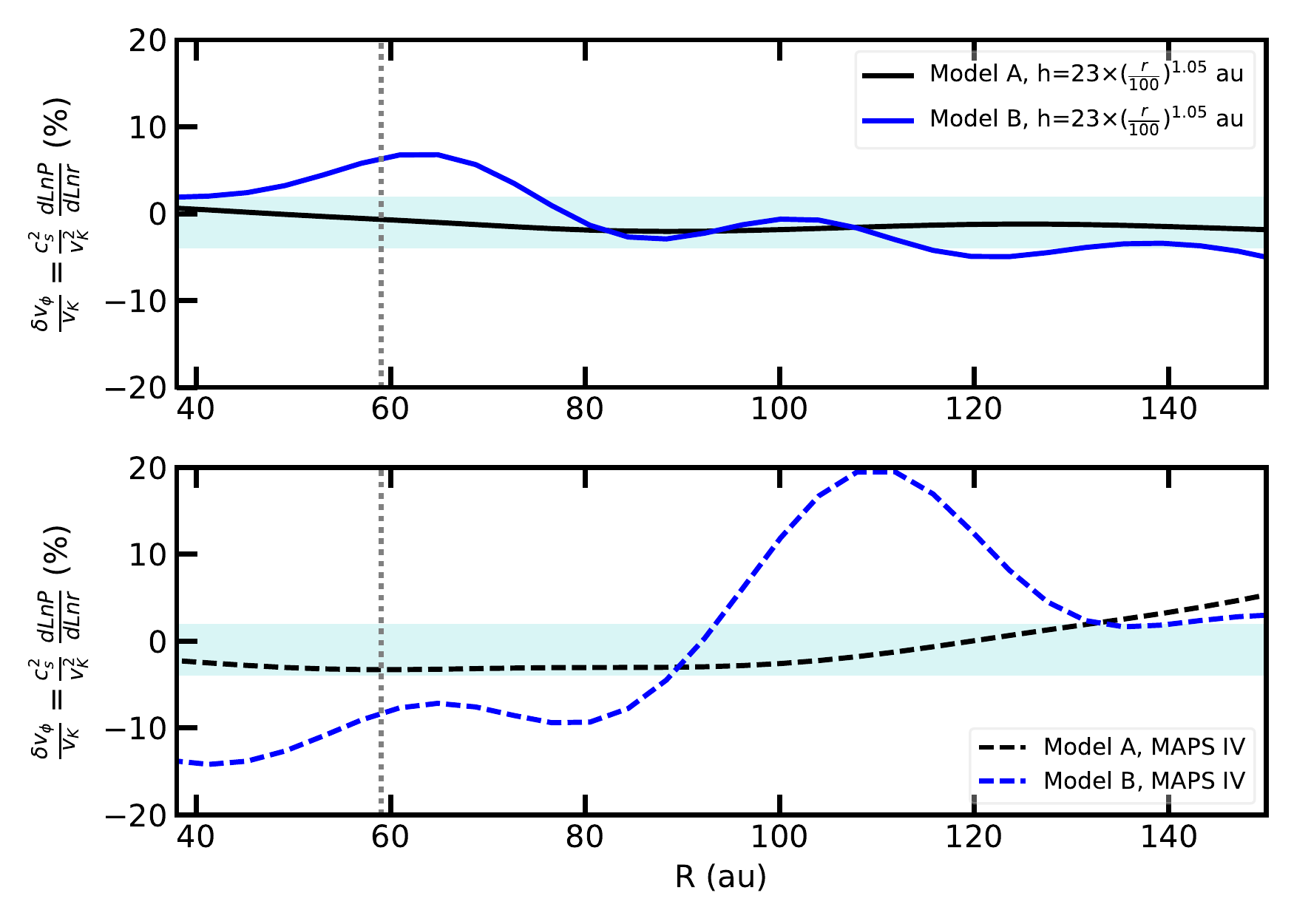}
    \caption{Expected velocity deviations in our models  inferred from the pressure profiles.  We inferred the velocity deviations using the relationships described by \cite{Rosotti..et..al}. The shaded area illustrates the range of kinematic deviations from the data (see Appendix \ref{app:kine}). In Model A the deviations are significantly smaller than in Model B. Model A does not present strong kinematic deviations from its pressure profile, while Model B  with a gas gap shows deviations of the order of 10\% in both surfaces.}
    \label{fig:kine_models}
\end{figure}

The MAPS program has obtained higher-resolution and higher-S/N data than used by \citet{Teague..et..al..2018} and we can revisit this question with independent data.   The deviations from the Keplerian field as derived from MAPS CO J=2--1 data are shown in Appendix \ref{app:kine} (Fig.~\ref{fig:kinematics data}) and are of order 2\% in a relative sense ($\delta v/v$). 
In Fig.~\ref{fig:kine_models} we illustrate the expected kinematic deviations in our models on this same scale  using the relations for the gas structure from \citet{Rosotti..et..al} applied to our models. The Keplerian deviations in Model B range over 30\% (peak to peak) within the framework of the gap and are clearly inconsistent with the data that show only small $\sim 2$\% deviations.  Model A on the other hand does show some structure but is limited to deviations of order 8 \% at the gap edges. Fitting the kinematic deviations in the AS 209 disk goes beyond the scope of this paper, but a large drop in H$_2$ surface density of an order of magnitude would induce almost a factor of 10 larger than observed in the data. 

 Kinematical analysis within gas surface-density gap scenarios have also been explored by \cite{Rab..2020}; in this case for the HD 163296 disk. This is a different disk; however, their findings about kinematic deviations in gas-depleted gaps are consistent with our results for AS 209. Keplerian deviations in gas-depleted gaps are significant ,$\sim$ 10\%, compared to dust gaps without gas depletion. 
 Thus, analyzing the kinematic data in two independent data sets supports smoother transitions at the location of the CO column-density gap and the C$^{18}$O emission gap. A significant H$_2$ gas depletion causes more abrupt variations in both profiles that are not observed in the data.




Associating the gap in CO emission with gas depletion also represents a disconnect with the planetary mass  required  to carve the gaps in the dust and gas mass distribution. To reach a $\sim$ 50\% CO depletion in AS 209 through a decrease in the H$_2$ surface density \citep{zhang20_maps}, a 0.2 $M_{\mathrm{Jup}}$ planet would be required. Such a planet carving the dust gaps would also need to perturb the gas in a much wider region encompassing the two dust gaps without a common center. Thus, the disconnect between the dust and the gas, in terms of the mechanisms for gap opening, invalidates a general H$_2$ depletion in the disk, supporting the idea that the CO distribution in AS 209 is the result of a  chemical effect on CO.

We are not able to rule out gaps in the H$_2$ gas entirely, as there are small fluctuations in the observed pressure profile that hint at gas pressure differences corresponding to the gaps seen in the dust distribution.
However, any perturbation to the gas surface density must be rather shallow and less broad that what is observed for CO emission. 

\subsubsection{Testing Kinematic Deviations in Hydrostatic Equilibrium Models}

We also test whether or not our result, i.e., differences in the kinematic deviation in models with and without a H$_2$ density structure, stands in models when the hydrostatic equilibrium is considered. We look at the third iteration of a thermochemical model with hydrostatic equilibrium to see its effects on the kinematic structure. Doing a detailed analysis of the effect of hydrostatic equilibrium is not the goal of this work; a more complex and detailed analysis is required. Nevertheless, our test allows us to understand if accounting for the hydrostatic equilibrium changes our findings. We test our results by comparing the $^{12}$CO $J=2-1$ radial emission profiles and estimated temperature of the gas.  This represents a change from our previous practice of comparison to the CO column densities.  For this test we match the $^{12}$CO $J=2-1$ emission profile. We start our vertical hydrostatic equilibrium runs with the models matching the CO $J=2-1$ emission as they get us closer to a smoother solution without impactful changes in fewer iterations, while matching the CO emission.

\begin{figure}
    \centering
    \includegraphics[width=0.45\textwidth]{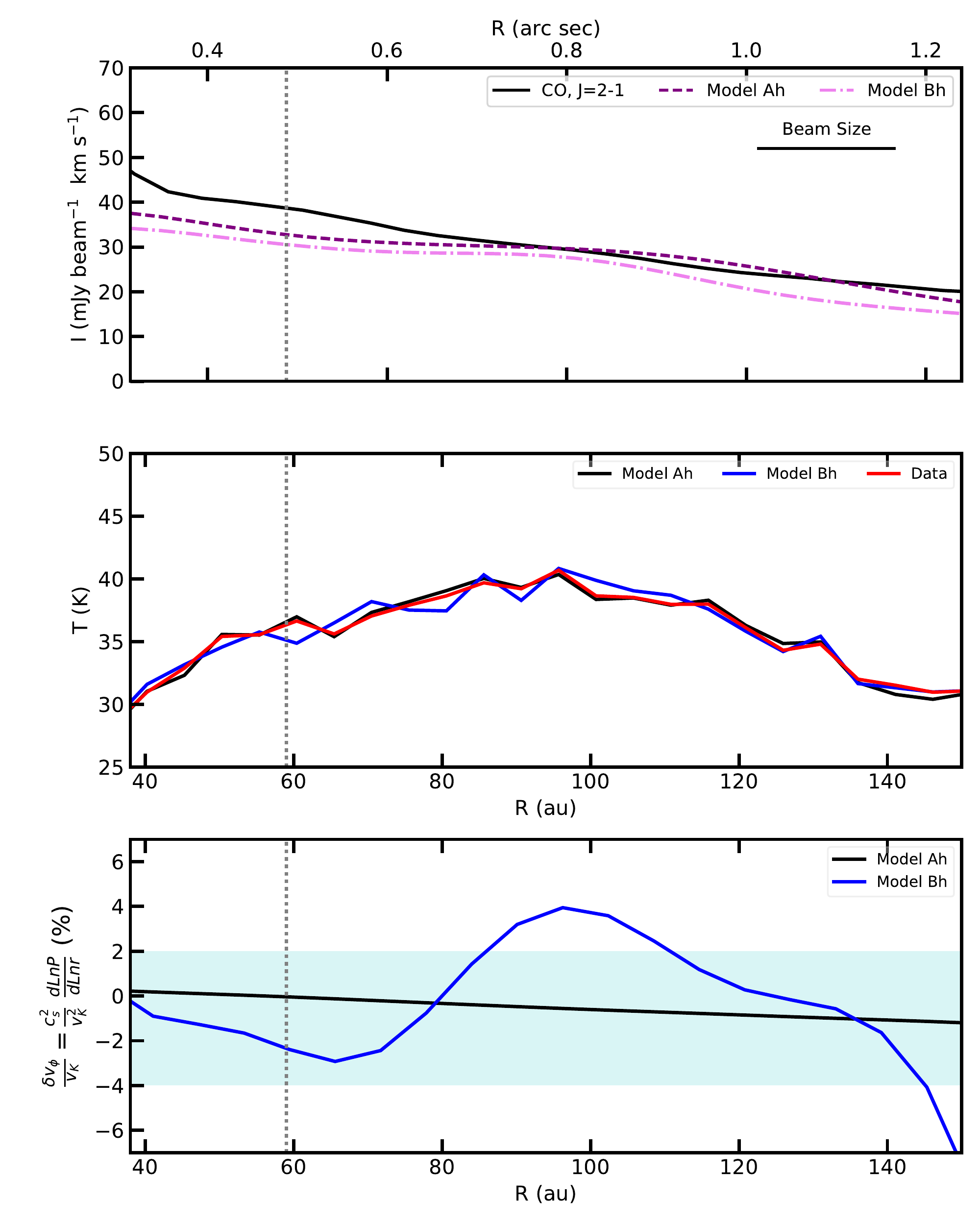}
    \caption{Comparison of data and expected values of thermochemical models with three iterations of hydrostatic equilibrium. \textbf{Top:} CO $J=2-1$ emission profile. \textbf{Middle:} thermal cuts for both models matching the brightness temperature of the CO $J=2-1$ line \citep{law20_rad}. \textbf{Bottom:} expected kinematic deviations in Models Ah and Bh with the range from  the observations represented by the shaded area. The models show that the CO $J=2-1$ is, to a small uncertainty, well traced in both cases. The kinematic deviations in Models Ah and Bh show that even when a few iterations of hydrostatic equilibrium are considered, the results stand, i.e., the kinematic deviations in Model Bh with a H$_2$ gas depletion are stronger  and cover  a larger range than the ones observed in the data.}
    \label{fig:hydro}
\end{figure}

We show the result of our test in Figure \ref{fig:hydro}. In order to reproduce the spatially resolved CO $J=2-1$ emission profile we had to increase the thickness of the disk with respect to \cite{zhang20_maps} to $h/r$=0.8 and the flaring index to $\psi$=1.35. Using those values we are able to match the CO emission in our region of interest. We refer to these models with hydrostatic equilibrium iterations as Model Ah and Model Bh. The temperature along the CO-emitting height in these models is within 10 K from the values inferred from the data \citep{law20_rad} in Models Ah and Bh. By slightly changing the thickness and flaring of the disk in the model to reproduce $^{12}$CO emission, these models find the warm molecular layer, the vertical zone where CO is not frozen onto grains or photodissociated, to be vertically higher in the disk. This also changes the overall CO abundance structure (see Appendix \ref{app:kine}) which may not fully reproduce the overall CO column density.  Nonetheless, we are able match the $^{12}$CO emission profiles even in models with hydrostatic equilibrium turned on.

To achieve consistency and uniformity between models we use the emitting layers for the hydrostatic test as the ones that fit the observed CO $J=2-1$ brightness temperature in the disk \citep{law20_rad} as shown in Fig.~\ref{fig:hydro} (middle panel).  Even with 10 K deviations in temperature, the major player in the pressure profile is the density along the emitting layer, which dominates any changes in the pressure profile.  Thus in models with a smooth density profile (Models A and Ah) we see small kinematical deviations, which is not the case for Models B and Bh where the H$_2$ surface density is assumed to contain a large gap.
We observe that the kinematic deviations are smoother due to the hydrostatic balance (e.g. compare kinematic residuals in Fig.~\ref{fig:kine_models} and Fig.~\ref{fig:hydro}); however, we still observe that the deviations in the model with H$_2$ gas depletion, while having values $\sim$ 4 \%, cover a range of the order of 7\%-8\% peak to peak, which is stronger than that seen in the CO kinematical analysis of  $\sim$ 2\% at most ($>$3.5 $\sigma$). Thus, we can conclude that our results stand in our test with hydrostatic equilibrium models, i.e., CO depletion is the main cause of the C$^{18}$O $J=2-1$ emission gap in AS 209.

\section{C$_2$H as a Tracer of Active Chemistry} \label{sec:C2H}

\subsection{C$_2$H column densities}

\begin{figure}
    \centering
    \includegraphics[width=0.5\textwidth]{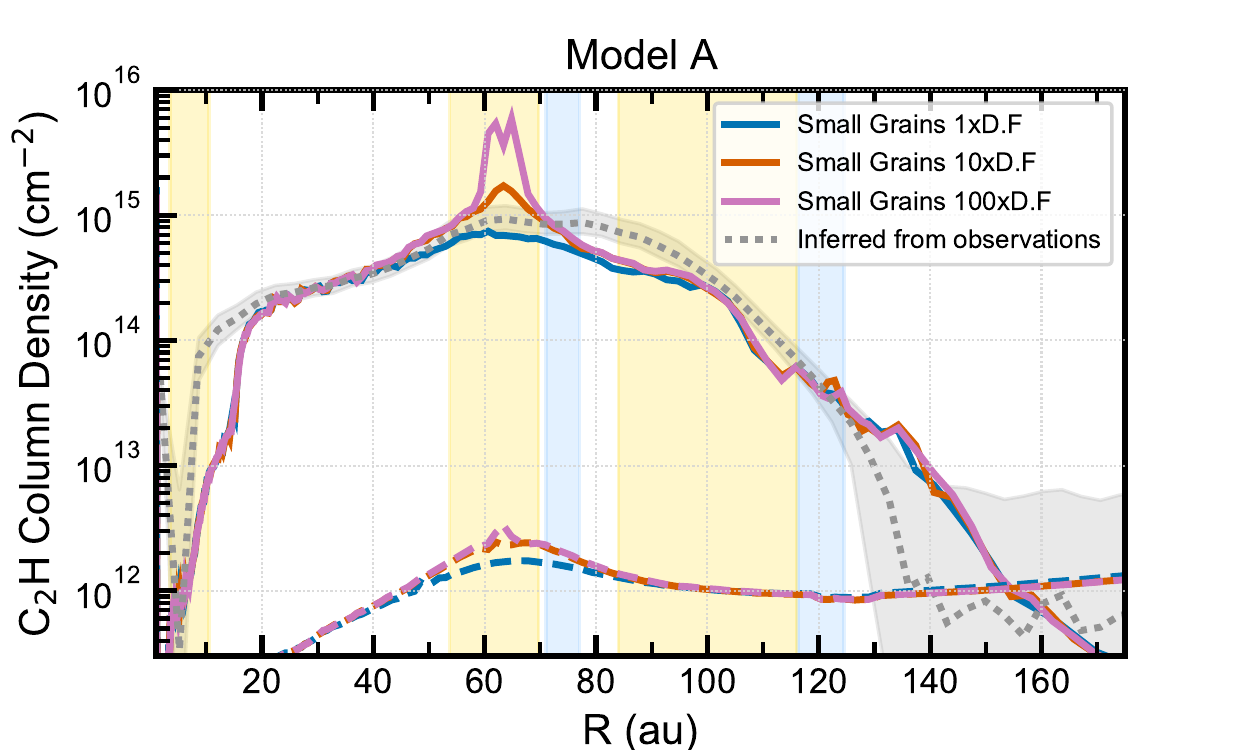}
    \includegraphics[width=0.5\textwidth]{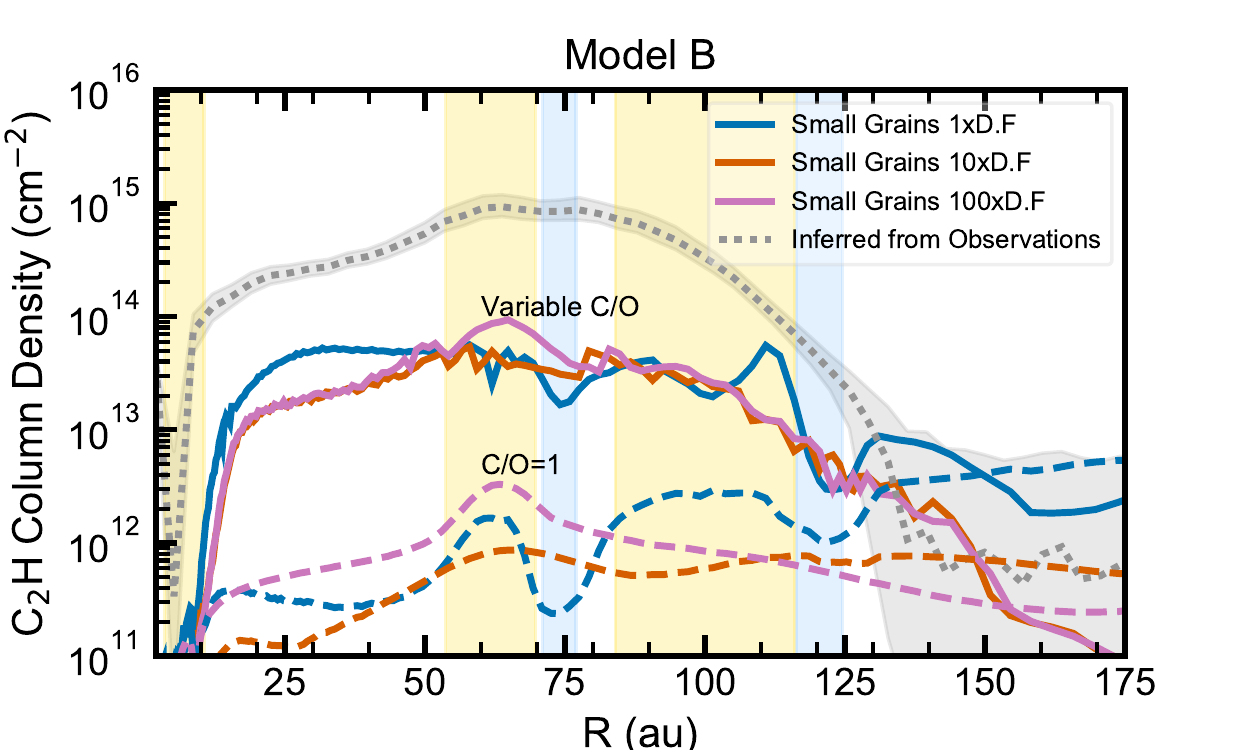}
    \caption{C$_2$H column-density profiles compared with the inferred values from observations. \textbf{Top:} simulation set for Model A. \textbf{Bottom:} simulation Set for Model B. The shaded areas in yellow and blue are the dust gaps and dust ring, respectively. Variable C/O models are shown in continuous lines, while the dashed lines are the models with a constant C/O ratio equal to 1. Our simulations show that depleting the number of small dust grains increases the C$_2$H column density by a factor of a few. However the main player in the C$_2$H column density is the C/O ratio. }
    \label{fig:C2H_ncols}
\end{figure}

If the abundance of CO is being reduced in planetary gaps due to changes in the physical conditions of the disk \citep{Alarcon..et..al..2020}, we expect that other species might also be affected as CO is the main carrier of carbon and oxygen in the disk gas.
 So, we look at possible fingerprints of CO processing using  C$_2$H, as it has bright emission lines that are commonly observed in protoplanetary disks \citep{Kastner14,Bergner..et..al..2020,Miotello..et..al..2019}. In Figure \ref{fig:C2H_ncols} we compare the C$_2$H column densities in both Models A and B with the retrieved values in \cite{guzman20} for AS 209.  In Model A, we show solutions with C/O = 1 and C/O = 2 between 20 and 110 au to demonstrate that an elevated C/O ratio is a requirement to increase the C$_2$H production matching the observations, with a subsequent reset at $r\sim$ 110 au to C/O = 0.4.  This result is reproduced in Model B as well. Thus, the C/O ratio is more important in determining the C$_2$H column density than  the actual surface-density profile, agreeing with the results in \cite{bosman20_CtoO} for the case of a smooth disk. Model A has a larger C$_2$H column density compared to Model B for the elevated C/O = 2 ratio profile, while that trend is not clear when C/O = 1.  C$_2$H production depends on the local density and UV field \citep{Bosman..2021}. Thus, the C$_2$H gas is localized in a narrow layer in the disk (see Figure \ref{fig:Model_C2H_set} tracing the local condition in that region. Therefore, C$_2$H is not a reliable radial surface-density tracer as it does not show a consistent correlation between Models A and B when the C/O ratio or the small-grain abundance is changed. Our models also assumed that the C/O ratio does not have vertical variations at a given radius. It is possible that the same C$_2$H can be reproduced by vertically localized increments in the C/O ratio without changes in the H$_2$ surface density. Such increases are consistent with results from numerical models \citep{Krijt..et..al..2020}. An elevated C/O ratio dependence for C$_2$H insensitive to surface-density changes is also consistent with the findings of \cite{Miotello..et..al..2019} that  C$_2$H flux does not strongly correlate with the disk or dust mass.   We note that even with a high C/O ratio and large depletion in the abundance of small grains, we still underestimate the C$_2$H column density in Model B.
 
The C$_2$H column-density peak seems to require a radial trend in the C/O ratio  (Figure \ref{fig:C_to_Oratio}), with a higher C/O ratio where there is a higher C$_2$H column density. 
We note that these increases in the C/O ratio occur in the same disk regions where the CO abundance is being lowered. Even though both effects may be related, we are not able to state a clear link between them as there are other possible chemical processes producing the radial variation in the C/O ratio.

\section{Discussion} \label{sec: Discussion}

\subsection{planet--disk Interaction vs Chemical Processing}


The CO column-density gap morphology, i.e., width, depth, and location, allows us to put an upper bound on the mass of the planet by using the relationships described by \citet{Kanagawa..et..al..2016,Kanagawa..et..al..2017}. Any planet more massive than this upper bound would carve a gas gap deeper than the CO drop in column density. Using the values from \cite{zhang20_maps} for the gap in CO column density; the gap width, $w_{\rm gap} = 13$ au, the gap location $R_{\rm gap} = 59$ au and the depletion $\delta_{\rm gap} = 0.47$ with an $\alpha$-viscosity = 10$^{-3}$, the mass of the gap-opening planet would be that of a sub-Jovian planet with a mass of 0.2 $M_{\rm Jup}$, thus this is still consistent with the values reported by \cite{Fedele..et..al..2018}, \cite{Zhang..DSHARP} and \cite{Favre..et..al..2019}. Therefore, CO being chemically processed does not disagree with the presence of a sub-Jovian planet carving the dust gaps in AS 209, similar to the scenario proposed by \cite{Dong..et..al..2017}.


\subsection{Carbon and CO Chemistry in Planet-forming Disks}
\label{sec:iso}
There are some significant CO chemical processes that support the CO depletion scenario discussed here. One such process that is not explicitly included in our modeling is CO isotopologue-selective photodissociation. \cite{Miotello..et..al..2014} showed that, particularly for the case of C$^{18}$O, isotopologue-selective photodissociation would decrease the inferred CO column density and the line emission as well. However, the CO column-density profiles inferred by \cite{zhang20_maps} using $^{13}$CO and C$^{18}$O line emission independently are in agreement, showing that isotopologue-selective photodissociation is not significant in AS 209. Even if its effect were to be significant, a strong isotopologue-selective photodissociation effect supports the scenario in which the emission gap in C$^{18}$O is explained by changes in the abundance of  C$^{18}$O rather than a strong gas depletion caused by a massive planet.


  Based on our mass-independent analysis  of the C$^{18}$O  emission structure using the pressure profile, we find that the abundance of CO is lowered locally in the AS 209 disk. The localized abundance depletion of CO in AS 209 reduces the overall oxygen in the system, likely enabling C$_2$H production \citep{Bosman..2021}.  Water ice is likely frozen and locked in the midplane \citep[][]{Hogerheijde11,Du17} and it is CO that is the main carrier of oxygen in the outer disk beyond the CO$_2$ ice line at $\sim$ 5 au. However, our results also require extra available carbon. This appears to be a common result as \cite{bosman20_CtoO} show that at least three of the MAPS disks require C/O$>$ 1 and a depletion of the small-grains abundance.
The origin of this excess carbon is a matter of debate.  Carbon grain destruction might provide the source term \citep[][]{Anderson17, Gail17, Klarmann18}; this would be consistent with a reduction in small-grains optical depth and extra gaseous carbon. Alternately, it might come from a two-step process where CO is destroyed via ubiquitous He$^+$ ions followed by water-ice formation removing oxygen from the gas phase \citep[][]{Bergin14, Reboussin15, Yu..et..al..2017, Kamber..et..al..2019}. Our model is not a self-consistent exploration of these processes and we cannot distinguish between these two scenarios.  Whether the chemical processing of CO leads to abundance depletion or it is anticorrelated with the production of C$_2$H is uncertain and requires detailed models that couple dust growth to chemistry, such as those shown by \citet{Krijt..et..al..2020}.

From an observational perspective, a previous survey of a small sample of circumstellar disks did not show any anticorrelation between the integrated $^{13}$CO flux and C$_2$H luminosity \citep{Miotello..et..al..2019}. \cite{Bergner..et..al..2020} also showed that there is no global anticorrelation between C$^{18}$O and C$_2$H column densities, which is consistent with the behavior of the other MAPS sources \citep{law20_rad,guzman20}.  We attribute this lack of correlation to the radial structure of the CO abundance, hiding this chemical processing in unresolved observations. C$_2$H becomes abundant in narrower regions in the disk, while CO is present throughout the disk. 

\begin{figure}
    \centering
    \includegraphics[width=0.5\textwidth]{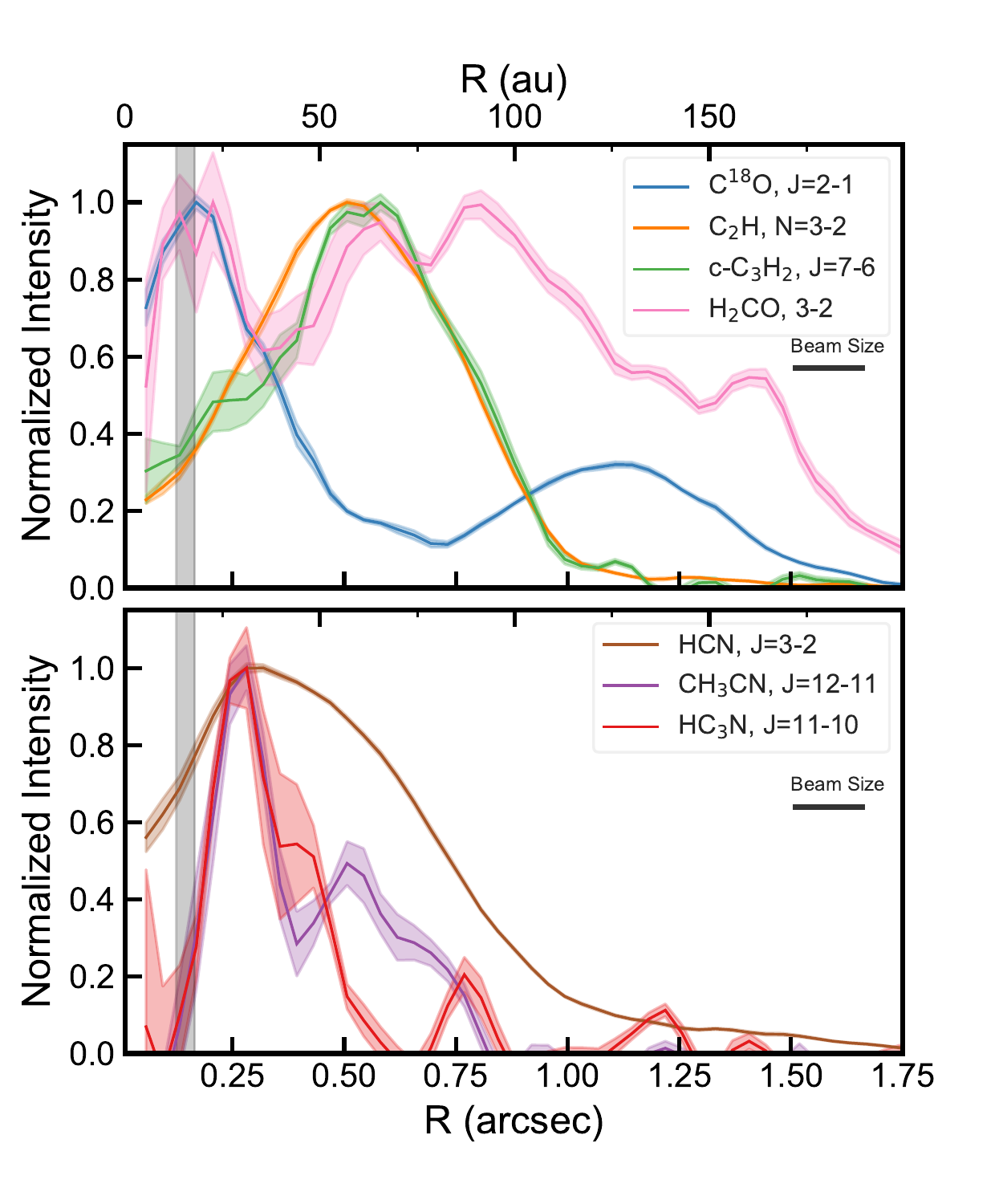}
    \caption{Normalized line emission of select species. \textbf{Top:} radial profiles of hydrocarbons and CO. We observe a multiple peak structure within the gas gap. \textbf{Bottom:} radial profiles of nitriles in AS 209. They peak close to 25 au, one of the inner continuum-emission rings. The black shaded area represents the CO ice line in the midplane.}
    \label{fig:Several_molecules}
\end{figure}

We also show the normalized radial intensity profiles of other carbon tracers in the upper panel of Figure \ref{fig:Several_molecules} \citep{law20_rad}. While C$^{18}$O exhibits a broad emission gap, other carbon tracers such as c-C$_3$H$_2$ and H$_2$CO have emission rings that coincide with that of C$_2$H.  Thus the abundance reduction of CO powers the production of these organic compounds. Moreover, C$_2$H and c-C$_3$H$_2$ show a broad and coherent ring and have similar formation paths \citep{Henning..et..al..2010}, while H$_2$CO has multiple structures.  H$_2$CO peaks just outside the inferred midplane CO ice line, while H$_2$CO also has additional structure just  beyond the C$_2$H ring. The outermost gap for H$_2$CO at $\sim$ 150 au might be associated with the second CO ice line located near the edge of the pebble disk \citep[][]{Cleeves16}.  

Another important chemical tracer of the local gas-to-dust ratio in the disk is DCO$^{+}$. \cite{Smirnov..et..al..2020} show that a  dust-poor but gas-rich gap produces more DCO$^{+}$ and HCO$^{+}$, increasing their column density by several orders of magnitude. Observing these tracers at high spatial resolution, i.e., resolving the dust gaps, would provide extra support to disentangle gas- and CO-depleted scenarios required to explain the CO emission in dust gaps. From our current analysis, we expect strong HCO$^{+}$ and DCO$^{+}$ emission in the CO column-density gap, which is observed in \cite{Favre..et..al..2019} and \cite{Aikawa20}.

The structure in the inner disk ($<$0$\farcs$4) is intriguing as this would be chemistry associated within the inner tens of astronomical units of this system. The bottom panel in Figure \ref{fig:Several_molecules} shows the distribution of some nitriles in AS 209. It is clear the emission from nitriles and CO falls off at $\sim$ 50 au while other species such as C$_2$H behave differently, starting to rise up after the inner 50 au. This distinct behavior in AS 209 points at a different chemistry between nitriles and hydrocarbons. The origin of this chemistry is uncertain.
This bright nitrile emission hints at an {\em active} chemistry associated with the inner disk that favors the production of these species in the disk's primary planet-forming region, with implications for the C/N ratio of compounds in the gas and within solids. 

\section{Summary} \label{sec: Summary}

We investigated the gas structure of AS 209 by analyzing high-resolution data from the MAPS ALMA Large Program \citep{oberg20}. These data show a broad depression in C$^{18}$O emission and a concurrent rise in C$_{2}$H emission around $\sim$ 60 au. We compared two possible scenarios to explain the observed C$^{18}$O emission gap: a local depletion in the gaseous H$_2$ surface density or a local CO abundance depletion. Both scenarios provided good fits of the observed CO column density and reproduced the $^{12}$CO and $^{13}$CO $J=2-1$ emission heights.   However, the pressure profile as estimated by local velocity deviations from the Keplerian flow by \citet{Teague..et..al..2018} and our own reanalysis of the kinematical deviations from the Keplerian flow in independent MAPS CO data, shows that the presence of a broad (13 au) order-of-magnitude deep  gas gap is not detected.  Thus we conclude that the CO abundance in AS 209 is locally reduced within a nearly smooth H$_2$ density profile and that chemical processing is active in this system. Based on the lack of significant structure in the overall gas density distribution, the presence of a Jovian-mass planet($>$0.2 $M_{Jup}$) in AS 209 is disfavored for $\alpha \leq 10^{-3}$. We cannot rule out the presence of a less-massive planet, which would be consistent with previous modeling of the CO emission substructure on the source \citep{Favre..et..al..2019}, although at a lower spatial resolution.

We also show that the chemistry that forms C$_2$H is insensitive to H$_2$ gas depletion, but in the AS 209 disk it appears to be correlated with the local reduction of the CO abundance. We speculate that this is due to the impact of the oxygen carried by CO which, if present, would hinder C$_2$H production by maintaining a C/O ratio lower than 1. Moreover, the processing associated with CO depletion can potentially produce C/O ratios that exceed unity. Beyond this association, there appears to be a rich and active chemistry associated with the early stages of planet formation in this system that remains to be understood.

\acknowledgments

This paper makes use of the following ALMA data: ADS/JAO.ALMA\#2018.1.01055.L. ALMA is a partnership of ESO (representing its member states), NSF (USA) and NINS (Japan), together with NRC (Canada), MOST and ASIAA (Taiwan), and KASI (Republic of Korea), in cooperation with the Republic of Chile. The Joint ALMA Observatory is operated by ESO, AUI/NRAO and NAOJ. The National Radio Astronomy Observatory is a facility of the National Science Foundation operated under cooperative agreement by Associated Universities, Inc.

We thank the anonymous referee for the feedback and constructive comment on this paper. F.A, A.D.B. and E.A.B. acknowledge support from NSF AAG Grant \#1907653. K.Z. acknowledges the support of the Office of the Vice Chancellor for Research and Graduate Education at the University of Wisconsin–Madison with funding from the Wisconsin Alumni Research Foundation, and support of NASA through Hubble Fellowship grant HST-HF2-51401.001. awarded by the Space Telescope Science Institute, which is operated by the Association of Universities for Research in Astronomy, Inc., for NASA, under contract NAS5-26555. J.B. acknowledges support by NASA through the NASA Hubble Fellowship grant \#HST-HF2-51427.001-A awarded  by  the  Space  Telescope  Science  Institute,  which  is  operated  by  the  Association  of  Universities  for  Research  in  Astronomy, Incorporated, under NASA contract NAS5-26555.  R.T. and F.L acknowledge support from the Smithsonian Institution as a Submillimeter Array (SMA) Fellow. K.R.S. acknowledges the support of NASA through Hubble Fellowship Program grant HST-HF2-51419.001, awarded by the Space Telescope Science Institute, which is operated by the Association of Universities for Research in Astronomy, Inc., for NASA, under contract NAS5-26555. J.K.C. acknowledges support from the National Science Foundation Graduate Research Fellowship under grant No. DGE 1256260 and the National Aeronautics and Space Administration FINESST grant, under grant no. 80NSSC19K1534.  Y.A. acknowledges support by NAOJ ALMA Scientific Research Grant CODE 2019-13B and Grant-in-Aid for Scientific Research (S) 18H05222, AND gRANT-IN-aID FOR transformative Reseacrh Areas (A) 20H05844 and 20H05847. J.D.I. acknowledges support from the Science and Technology Facilities Council of the United Kingdom (STFC) under ST/T000287/1. K.I.\"O. acknowledges support from the Simons Foundation (SCOL \#321183) and an NSF AAG Grant (\#1907653).  I.C. was supported by NASA through the NASA Hubble Fellowship grant HST-HF2-51405.001-A awarded by the Space Telescope Science Institute, which is operated by the Association of Universities for Research in Astronomy, Inc., for NASA, under contract NAS5-26555. C.J.L. acknowledges funding from the National Science Foundation Graduate Research Fellowship under Grant DGE1745303. C.W. acknowledges financial support from the University of Leeds, STFC and UKRI (grant Nos. ST/R000549/1, ST/T000287/1, MR/T040726/1). M.L.R.H. acknowledges support from the Michigan Society of Fellows. R.L.G. acknowledges support from a CNES fellowship grant. F.M. acknowledges support from ANR of France under contract ANR-16-CE31-0013 (Planet-Forming-Disks)  and ANR-15-IDEX-02 (through CDP ``Origins of Life"). G.C. is supported by NAOJ ALMA Scientific Research
grant code  2019-13B. S.M.A. and J.H. acknowledge funding support from the National Aeronautics and Space Administration under grant No. 17-XRP17 2-0012 issued through the Exoplanets Research Program.  J.H. acknowledges support for this work provided by NASA through the NASA Hubble Fellowship grant \#HST-HF2-51460.001-A awarded by the Space Telescope Science Institute, which is operated by the Association of Universities for Research in Astronomy, Inc., for NASA, under contract NAS5-26555. Y.L acknowledges the financial support by the Natural Science Foundation of China (Grant No. 11973090).

%

\vspace{5mm}

\software{\texttt{RADMC3D} \citep{RADMC3D}, \texttt{RAC2D} \citep{Fujun..Ted..RAC2D}, astropy \citep{AstropyI,AstropyII} }

\appendix \label{appendix}

\section{AS 209 as a planet formation laboratory} \label{sec: AS209}

\begin{figure}[hb!]
    \centering
    \includegraphics[width=0.49\textwidth]{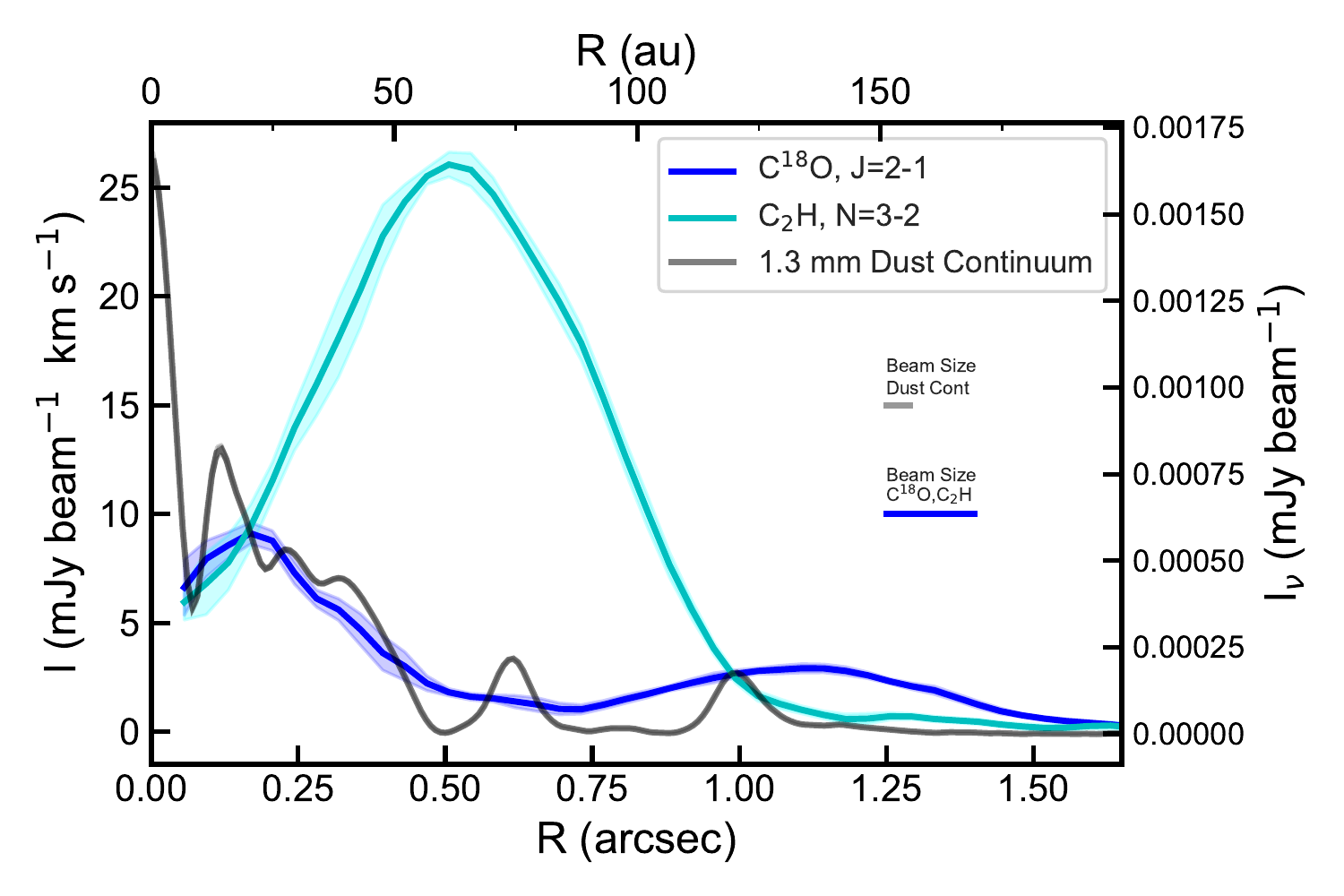}
    \includegraphics[width=0.49\textwidth]{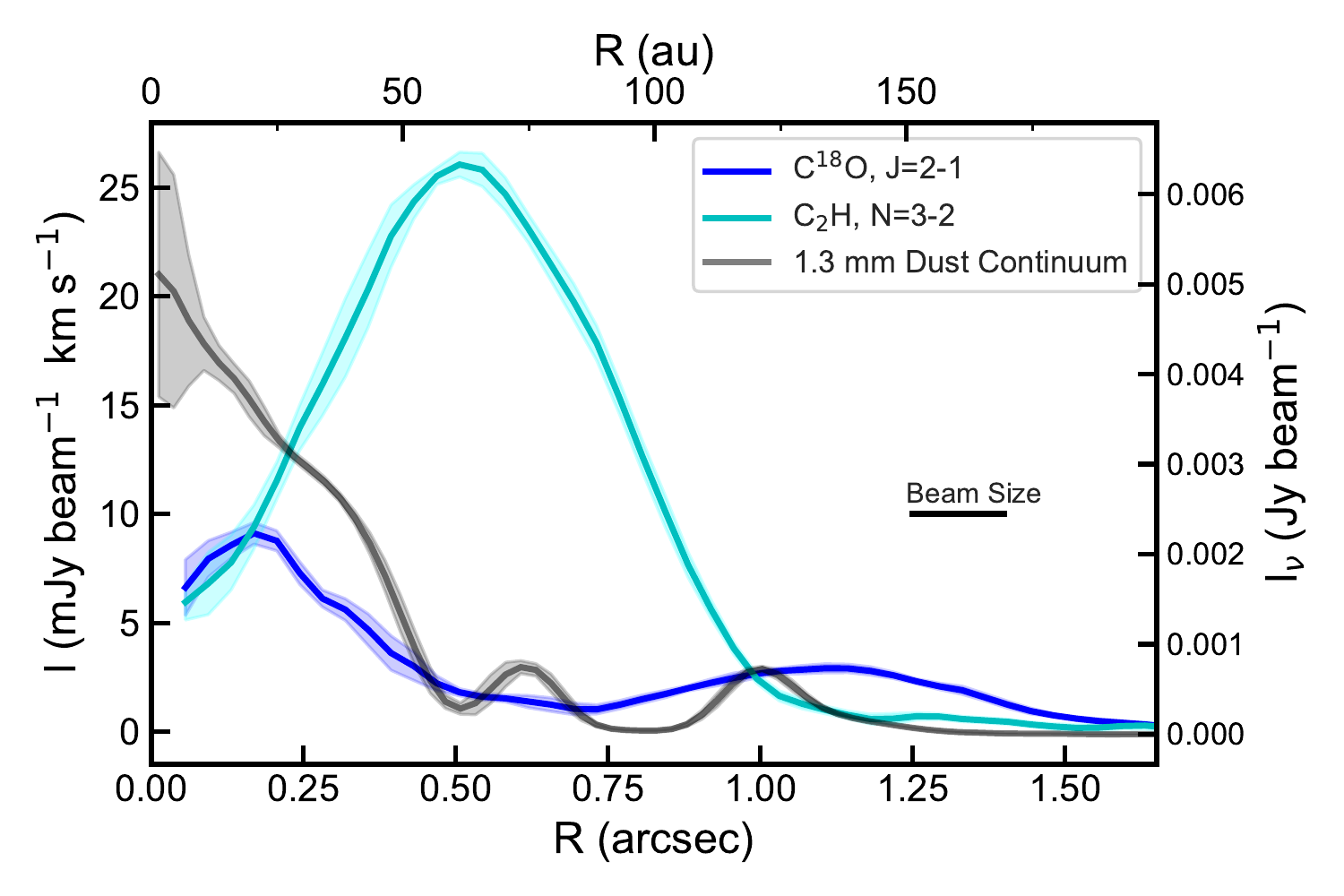}
    \caption{C$_2$H and C$^{18}$O radial profile for the velocity-integrated emission lines in ALMA Band 6, while the gray line displays the continuum-emission profile from DSHARP \citep{DSHARPI,Vivi..et..al2018} on the left and from MAPS \citep{oberg20} on the right. We observe that C$^{18}$O shows a wide gap before showing a second peak after the outer ring at 120 au.  C$_2$H shows an inverse correlation to C$^{18}$O, peaking at the  location of the C$^{18}$O gap, making AS 209 a unique source for the study of CO chemical processing among the sources in the MAPS survey.}
    \label{fig:CO_C2H_cont}
\end{figure}

In this work we focused on the C$^{18}$O emission gap at $\sim$80 au and the C$_{2}$H emission ring around the same location, while the dust continuum presents one ring at this location with two gaps at each side of it  (see Fig \ref{fig:CO_C2H_cont}). It is noteworthy that understanding the origin of substructures in line emission is more complex than in dust continuum. The dust substructures are linked to the dynamical evolution of the disk, while substructures in line emission can also be linked to the disk's chemical evolution.

Several studies have proposed the presence of one or more planets at different locations within AS 209 as the origin of the dust rings and gaps. \cite{Zhang..DSHARP} proposed that a single sub-Jovian planet with an $\alpha \sim$ 10$^{-4}$  (0.2 $M_{\rm Jup}$) is able to carve the multiple gaps in AS 209 \citep{Bae..Zhu..Lee..2017,Dong..et..al..2017}. With this planet located at 100 au, models suggest that the resulting interactions can explain the 100 au gap and all the interior dust substructure. \cite{Fedele..et..al..2018} explored the possibility of a second planet in the inner dust gap at 62 au. They found that the presence of a 0.05 $M_{\rm Jup}$ planet also fits the dust substructure, although such a planet is not necessary to explain the observations. \cite{Dong..et..al..2018} suggested that a 0.09 $M_{\rm Jup}$ planet embedded in the 80 au continuum ring could explain the location and depth of the dust gaps at each side of the ring. 

\cite{Favre..et..al..2019} were the first to constrain on  a possible planet scenario in AS 209 using gas emission from CO at a spatial resolution of $\sim$ 0$\farcs$25. They claimed the need of gas depletion to match the C$^{18}$O gap in the emission profile, which can be explained by a 0.2 M$_{\rm Jup}$ planet at the outer dust gap at $\sim$100 au. Independently, \cite{Teague..et..al..2018} find velocity perturbations around three locations in the disk associated with radial changes in the pressure profile from CO observations. One of these perturbations was associated with a wide gas gap at $\sim$ 50 au with 80\% gas depletion. 

\subsection{Comparison with DSHARP}

Figure \ref{fig:CO_C2H_cont} shows the radial-velocity-integrated intensity profile in C$^{18}$O $J=2-1$ and C$_2$H $N=3-2$ from the MAPS survey in ALMA Band 6, along with the dust continuum emission from MAPS and DSHARP at a higher spatial resolution \citep{DSHARPI}. The C$^{18}$O emission profile shows a wide emission depression centered at 88 au with a width of 47 au, while C$_2$H shows an emission ring centered at 70 au with a width of 68 au \citep{law20_rad}. The gap in C$^{18}$O emission translates into a decrease in the CO column density, being depleted by at least 30\% when compared with a smooth surface-density profile \citep{zhang20_maps}. Even at the MAPS spatial resolution (0$\farcs$15), we still observe that the gap in the C$^{18}$O $J=2-1$ emission is broader than the dust continuum-emission profile, showing that the structure is resolved.

\begin{figure}[hb!]
    \centering
    \includegraphics[width=0.6\textwidth]{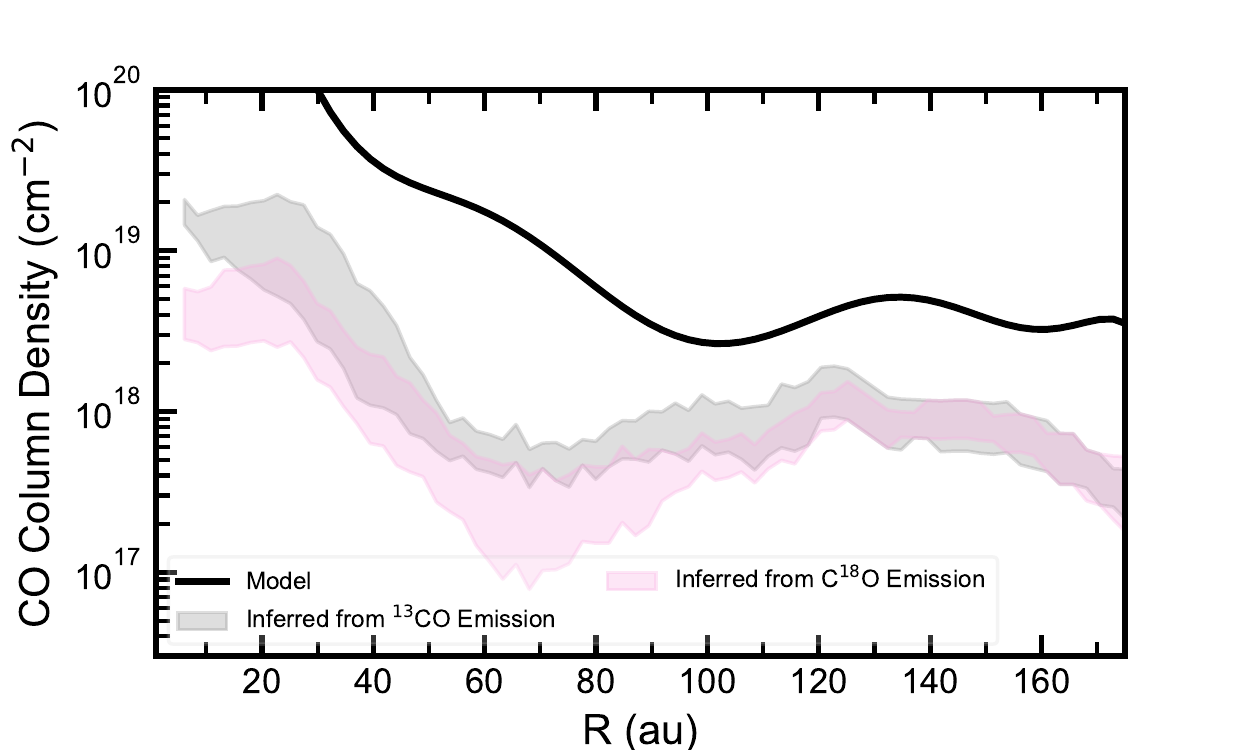}
    \caption{Comparison between the inferred CO column densities from \cite{zhang20_maps} and a model using the gas surface-density from the hydrodynamic simulations in DSHARP \citep{Zhang..DSHARP}. The gas surface density profile was scaled to the disk mass used in our simulations and with the same abundances listed in Table \ref{tab:Abundances} without any extra CO depletion. The profile using the the DSHARP gas surface density overestimates the CO column density by an order of magnitude, and the location of the dips and peaks do not match the ones observed in CO, supporting the CO-processing solution.}
    \label{fig:DSHARP_comp}
\end{figure}

We also ran a model using the gas surface-density profile matching the dust continuum emission from DSHARP using hydrodynamical simulations \citep{Zhang..DSHARP}. The gas surface density was scaled with the disk mass used in our simulations and the same abundances that were used in Model B (see Table \ref{tab:Abundances}).  We show the comparison in Figure
\ref{fig:DSHARP_comp} illustrating that CO depletion is needed to obtain the inferred CO column densities. Despite the hydrodynamical model in DSHARP fitting the dust continuum emission at a higher resolution, the dips and peaks in the CO column density do not match the ones in the profile from DSHARP, which supports the solution of dominant CO chemical processing rather than H$_2$ gas depletion.

\section{Models Setup and Structure}
\label{app:Model}

\subsection{Substructures}

The  gaps and rings were included following the same prescription as \cite{Alarcon..et..al..2020}, i.e., gaps and rings are a Gaussian modulation of the local surface density for a smooth parametric disk following the self-similar solution on an $\alpha$-viscosity disk \citep{Lynden-Bell_1974}.

Gaps have the following functional form:

        \begin{equation}\label{eq:gap}
    \Sigma(r) = \Sigma_{\rm basic}(1 - (1 - \delta_{\rm gap})\exp(-(r-r_{\rm gap})^2/2w_{\rm gap}^2)),
    \end{equation}

\noindent where $\delta_{\rm gap}$ is the depletion factor, $r_{\rm gap}$ the gap's location and $w_{\rm gap}$ the width of the gap.  Rings are enhancements of dust grains so we parameterize them with this equation:

  \begin{equation}\label{eq:ring}
            \Sigma(r) = \Sigma_{\rm dust}(1 + \delta_{\rm ring}\exp(-(r-r_{\rm ring})^2/2w_{\rm ring}^2)),
    \end{equation}
    
\noindent with  $\delta_{\rm ring}$  the enhancement factor, $r_{\rm ring}$ the ring's location, and $w_{\rm ring}$ the width of the ring.

\subsection{2D Thermal Structure Comparison} \label{app:2d T}

\begin{figure*}
    \centering
    \includegraphics[width=0.9\textwidth]{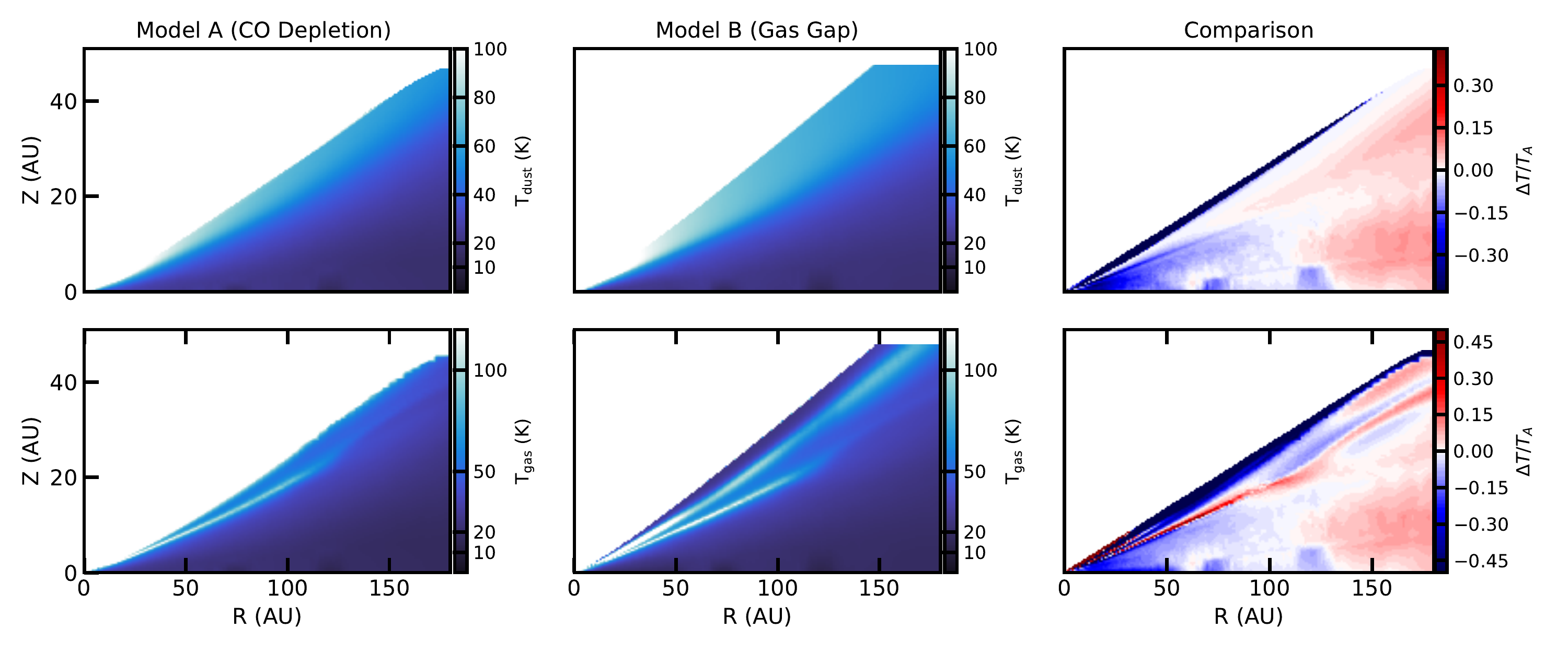}
    \caption{Thermal structure for Models A and B. \textbf{Top row:} dust temperature. \textbf{Bottom row:} gas temperature. The right column shows that the normalized residuals between models with $\Delta T = T_A - T_B$ is the difference between the temperature of Model A and Model B. The residuals are on the order of 10\%, which corresponds to 3--4 K degrees at the emission layer.}
    \label{fig:Model_T}
\end{figure*}

We show the 2D temperature field for each model in Figure \ref{fig:Model_T} and their respective difference. The differences in both cases are coming from the differences in the CO abundance between each model and small variations in the gas structure which are more significant in the gap. Nevertheless, the differences are usually less than 10\% between each model, so they may present some small radial or vertical variation for the abundance of given species, but it does not change the results of our models significantly. However, such differences between gas depletion and CO processing could potentially be traced in a deep analysis of the line-emission profiles of molecular tracers.

\subsection{2D Abundance Structure of CO and C$_2$H} \label{app:2d}

\begin{figure*}
    \centering
    \includegraphics[width=0.9\textwidth]{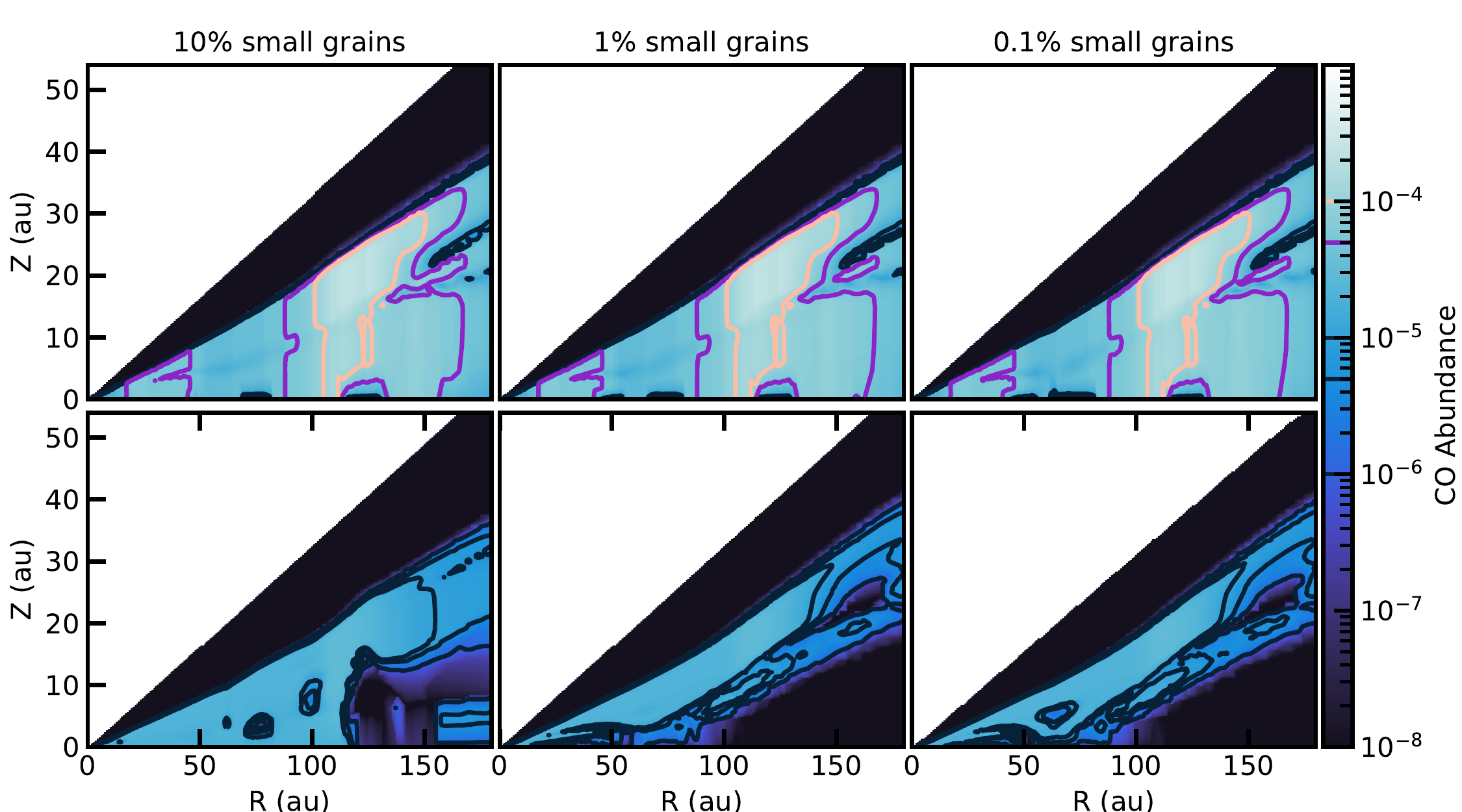}
    \caption{CO abundance structure for both model sets, A and B, with different localized depletion factors of small grains  from left to right.\textbf{Top row:} CO abundance structure for Model A, where CO has been depleted and the gas surface density has a smooth profile.\textbf{Bottom row:} CO abundance structure for Model B, where CO is considered to trace the gas surface density by a constant scaling or CO/H$_2$ ratio. In both models, there are not significant changes with different degrees of small-grains depletion. However, we observe that in Model A, CO has a more uniform vertical distribution than in Model B.}
    \label{fig:Model_CO_set}
\end{figure*}

We present the abundance structure in Figure \ref{fig:Model_CO_set} showing that in Model A, when CO has been depleted, the absolute CO abundance is lower than in Model B, which is compensated by a higher gas density. When we deplete CO, it has a more uniform abundance with height when compared to the second scenario. Therefore, even if in both scenarios the CO column-density radial profile is similar, the changes in the vertical distribution of CO could lead to changes in the emission radial profiles of CO isotopologues. By having different vertical structure, a detailed comparison of CO isotopologue emission profiles could provide differences between the emission heights of both models. Nevertheless, such approach goes beyond the scope of this paper.

\begin{figure*}
    \centering
    \includegraphics[width=0.9\textwidth]{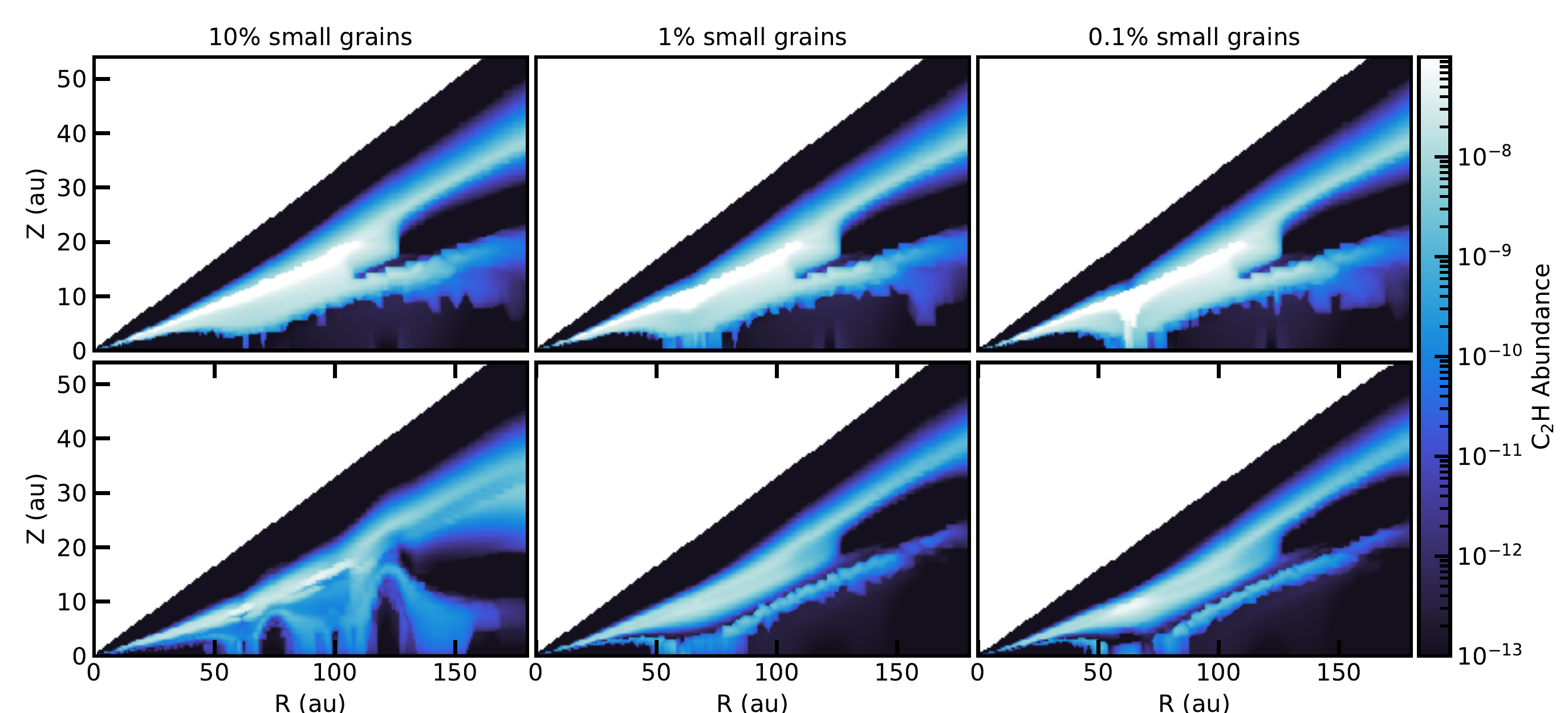}
    \caption{C$_2$H 2D abundance structure in each set of simulations.\textbf{Top row:} C$_2$H abundance structure for Model A.\textbf{Bottom row:} C$_2$H abundance structure for Model B. In the regions where the small dust is depleted, C$_2$H can be produced closer to the midplane. This causes the slight increases in C$_2$H column density observed in Fig. \ref{fig:C2H_ncols}.}
    \label{fig:Model_C2H_set}
\end{figure*}

If we compare the abundance structure of C$_2$H for each case (see Fig. \ref{fig:Model_C2H_set}), we observe that even though the vertically integrated column densities show changes of less than an order of magnitude, depleting the small grains causes the C$_2$H to be produced closer to the midplane, consistent in both Models A and B. Nevertheless, even though Models A and B produce C$_2$H at lower heights, the layers at which C$_2$H is being produced differ in the models. In Model B, there is more production of C$_2$H, i.e., higher C$_2$H abundances at lower heights. We link the higher C$_2$H abundances to a deeper UV penetration and a longer C$_2$H prevalence due to a lower gas density, i.e., fewer gas-phase reactions. 

\subsection{CO Emission Surfaces in AS 209} \label{app:surfaces}

\begin{figure*}
    \centering
    \includegraphics[width=0.9\textwidth]{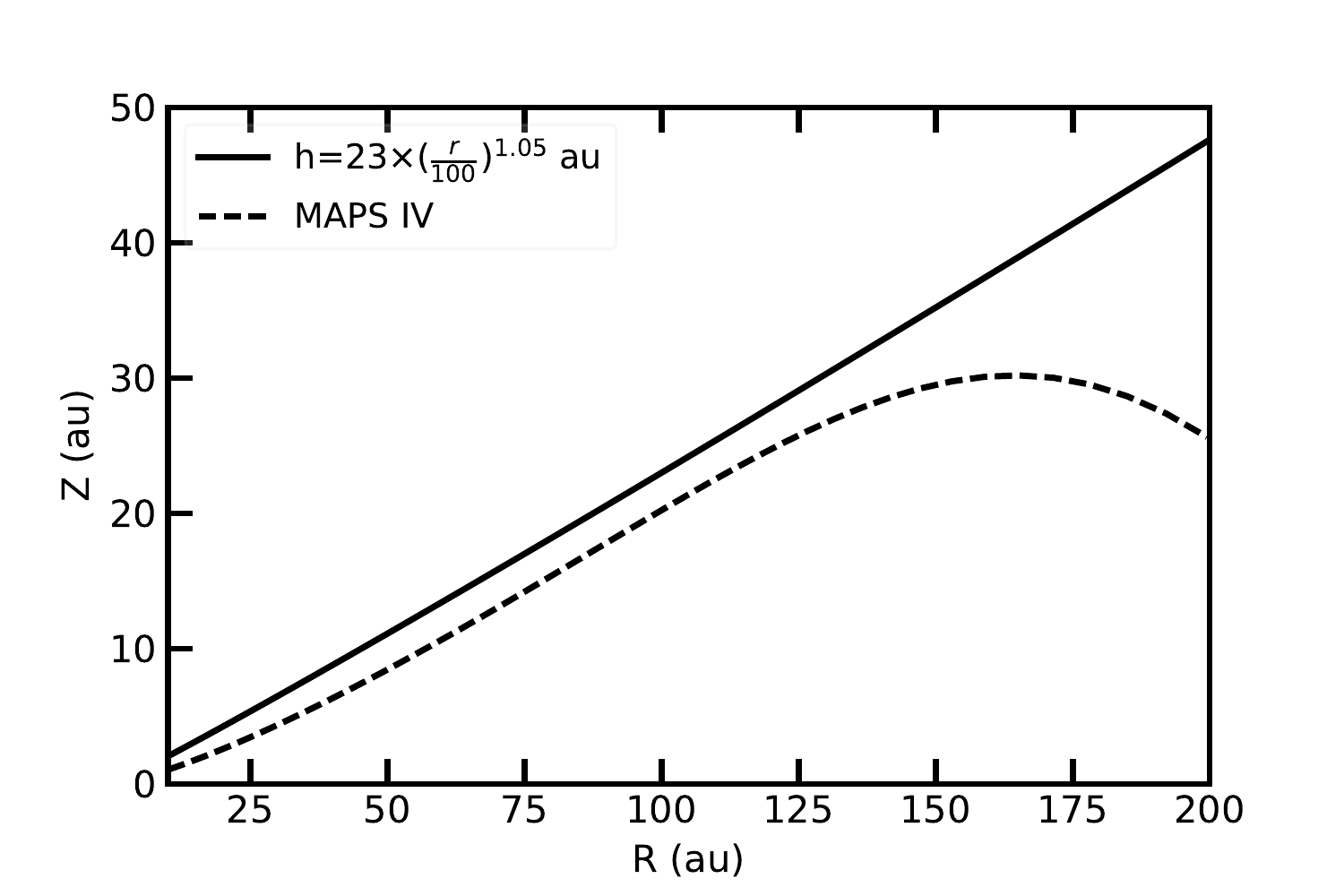}
    \caption{Emission heights found by \cite{Teague..et..al..2018} and \cite{law20_surf} for the CO $J=2-1$ emission coming from the AS 209 disk. These surfaces were used to extract the Keplerian deviation in our models and compare them with the observational ones.}
    \label{fig:HEIGHTS}
\end{figure*}

We show the difference between the CO emission layer for the $J=2-1$ transition in the AS 209 disk in Figure \ref{fig:HEIGHTS} from \cite{Teague..et..al..2018} and \cite{law20_surf}. There are subtle differences between each surface. The surface from \cite{Teague..et..al..2018} is slightly above the \cite{law20_surf} one, but they have a reasonable agreement considering uncertainties.

\section{Keplerian deviations in the AS 209 gas kinematics}
\label{app:kine}

\subsection{Keplerian deviations in the MAPS data}

\begin{figure*}
    \centering
    \includegraphics[width=0.7\textwidth]{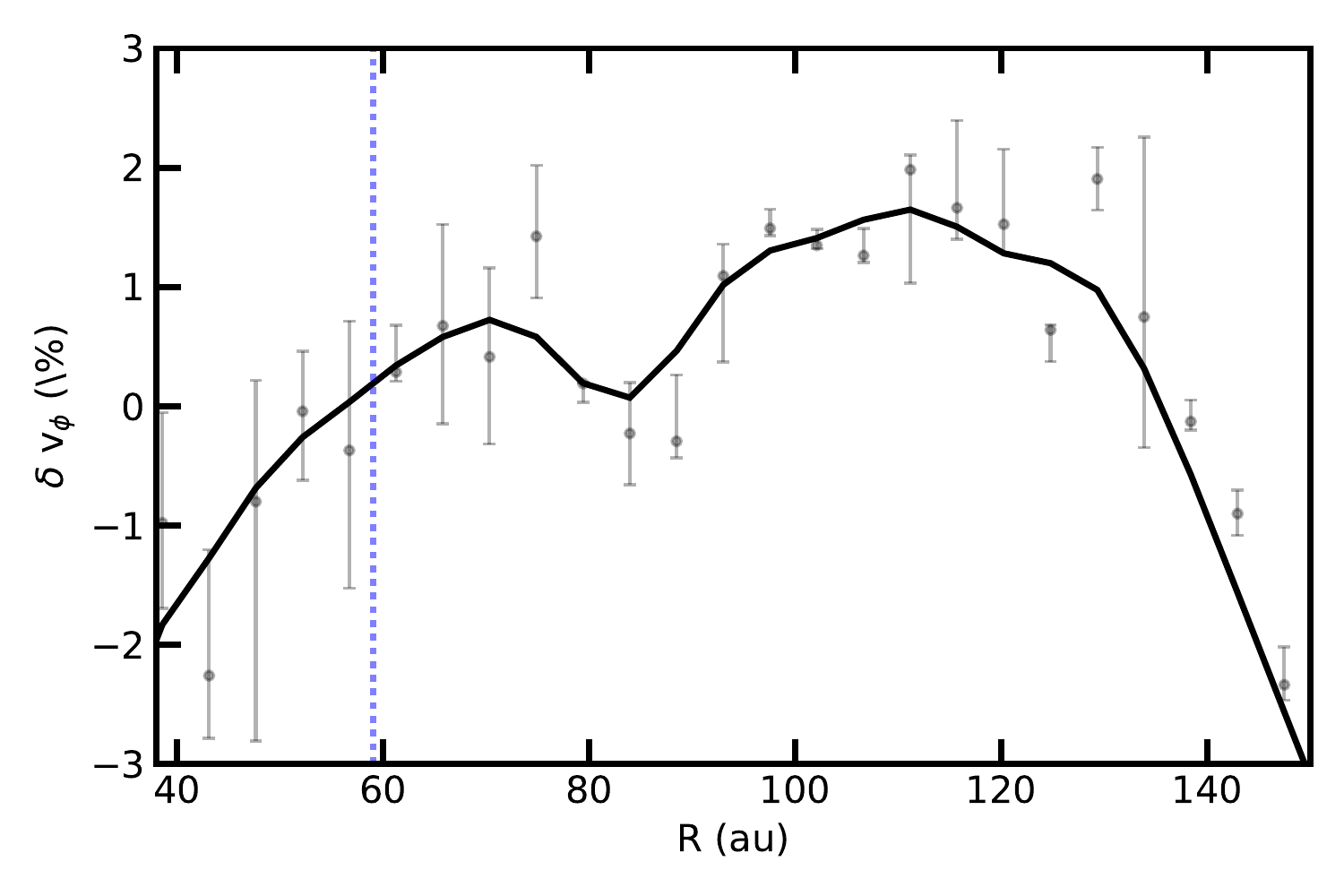}
    \caption{Kinematic deviations in the AS 209 disk using the CO data from MAPS \citep{oberg20}. The deviations show variations $<1$\% at the location of the CO column-density gap, which is illustrated by the vertical dotted line. The continuous profile is the expected profile through a convolution with a 0$\farcs$15 resolution ($\sim$ 18 au). The recovered deviations seem consistent with the profile obtained by \cite{Teague..et..al..2018} with lower-resolution data, reassuring that the kinematic deviation in AS 209 is not strong enough to cause a deep gas gap at 59 au.}
    \label{fig:kinematics data}
\end{figure*}

We show the kinematics deviations in the AS 209 disk using MAPS data at 0$\farcs$15 resolution in Figure \ref{fig:kinematics data} using methods described in \cite{Teague..et..al..2018}. The data shows that the inferred deviation in the disk is at the $\sim$ 1\% level in small spatial scales in the inner 100 au. Therefore, there is not a strong H$_2$ depletion in the disk at 59 au; otherwise, there should be a strong footprint in the kinematic deviations in the disk, as shown in Figure \ref{fig:kine_models}.

\subsection{Hydrostatic Equilibrium}

\begin{figure*}
    \centering
    \includegraphics[width=0.45\textwidth]{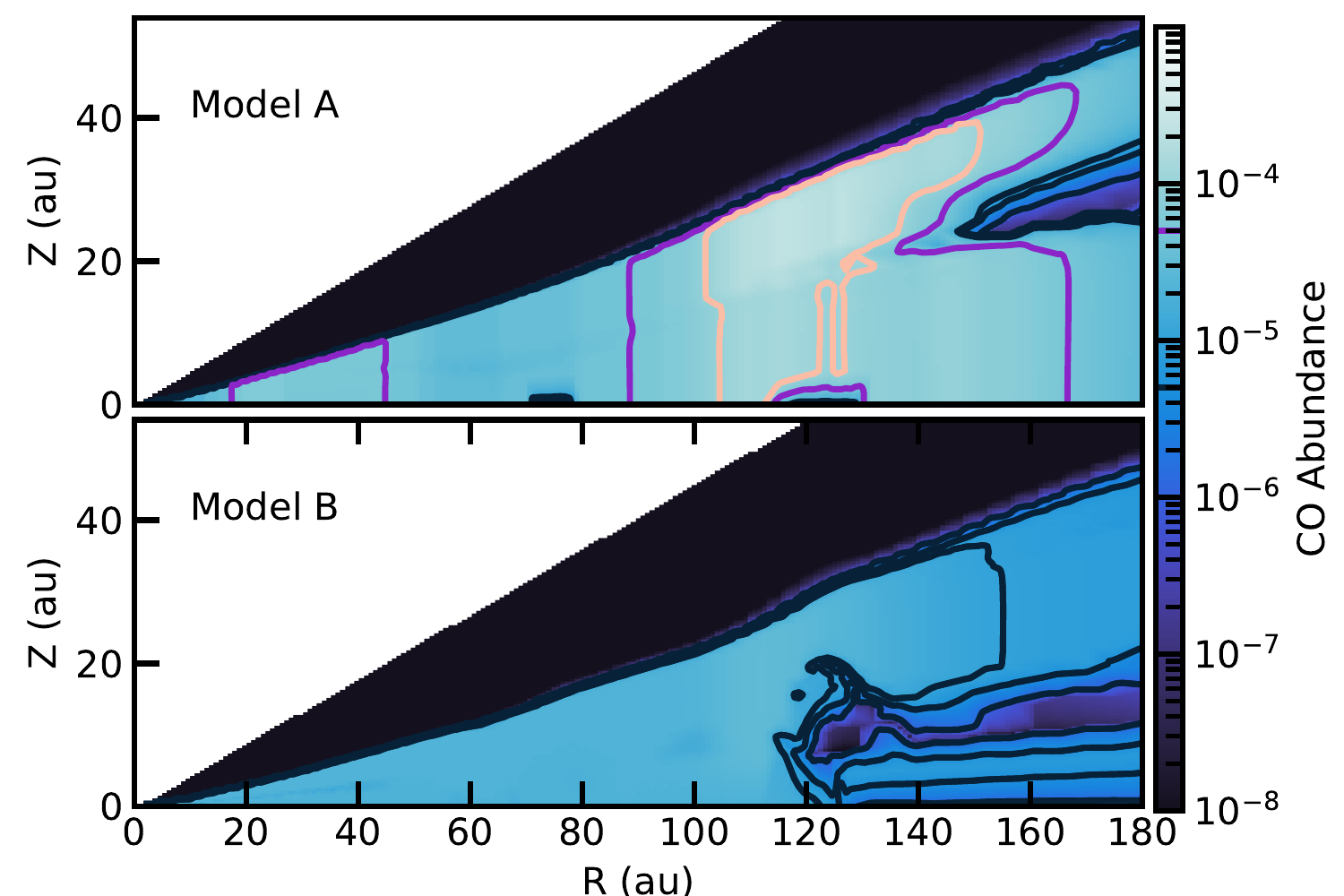}
    \includegraphics[width=0.45\textwidth]{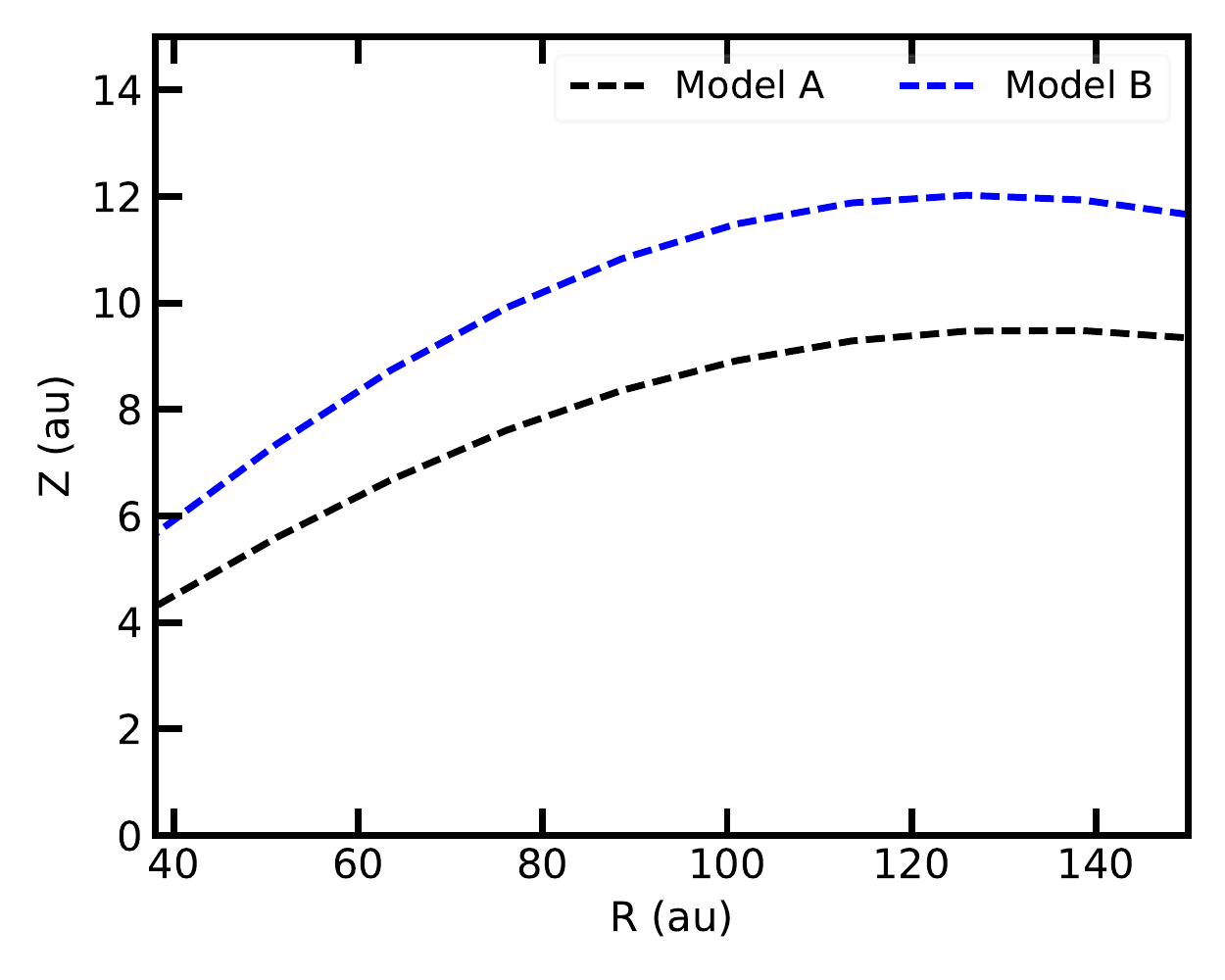}
    \caption{\textbf{Left:} 2D CO abundances of the model with hydrostatic equilibrium. \textbf{Right:} Layers at which we do the test for kinematic deviations.  The thicker and more flared disk geometry changes the CO abundance structure in the models with less freeze-out in the midplane beyond 120 au.}
    \label{fig:Hydro2}
\end{figure*}

We show the 2D fields of CO and the emitting heights in the hydrostatic equilibrium runs in Figure \ref{fig:Hydro2}. The emitting heights were taken at a different layer considering the thermal structure of the disk and the changes associated with the different geometry of the disk to match the CO $J=2-1$ line emission. Moreover, the CO 2D abundance structure shows changes behind the dusty ring at 120 au. Those changes are probably associated with the changes in the vertical distribution of the dust, changing the self-shielding in the disk beyond 120 au. Understanding the effect of hydrostatic equilibirum  requires a deeper and more sophisticated analysis that goes beyond the scope of this paper, but our test proves that our results remain in those conditions. 

\bibliography{MAPSVIII}{}

\begin{thebibliography}{}
\expandafter\ifx\csname natexlab\endcsname\relax\def\natexlab#1{#1}\fi
\providecommand{\url}[1]{\href{#1}{#1}}
\providecommand{\dodoi}[1]{doi:~\href{http://doi.org/#1}{\nolinkurl{#1}}}
\providecommand{\doeprint}[1]{\href{http://ascl.net/#1}{\nolinkurl{http://ascl.net/#1}}}
\providecommand{\doarXiv}[1]{\href{https://arxiv.org/abs/#1}{\nolinkurl{https://arxiv.org/abs/#1}}}

\bibitem[{{Aikawa} {et~al.}(2021){Aikawa}, {Cataldi}, {Yamato}, {Zhang},
  {Booth}, {Furuya}, {Andrews}, {Bae}, {Bergin}, {Bergner}, {Bosman},
  {Cleeves}, {Czekala}, {Guzm{\'a}n}, {Huang}, {Ilee}, {Law}, {Le Gal},
  {Loomis}, {M{\'e}nard}, {Nomura}, {{\"O}berg}, {Qi}, {Schwarz}, {Teague},
  {Tsukagoshi}, {Walsh}, \& {Wilner}}]{Aikawa20}
{Aikawa}, Y., {Cataldi}, G., {Yamato}, Y., {et~al.} 2021, arXiv e-prints,
  arXiv:2109.06419.
\newblock \doarXiv{2109.06419}

\bibitem[{{Alarc{\'o}n} {et~al.}(2020){Alarc{\'o}n}, {Teague}, {Zhang},
  {Bergin}, \& {Barraza-Alfaro}}]{Alarcon..et..al..2020}
{Alarc{\'o}n}, F., {Teague}, R., {Zhang}, K., {Bergin}, E.~A., \&
  {Barraza-Alfaro}, M. 2020, \apj, 905, 68, \dodoi{10.3847/1538-4357/abc1d6}

\bibitem[{{Anderson} {et~al.}(2017){Anderson}, {Bergin}, {Blake}, {Ciesla},
  {Visser}, \& {Lee}}]{Anderson17}
{Anderson}, D.~E., {Bergin}, E.~A., {Blake}, G.~A., {et~al.} 2017, \apj, 845,
  13, \dodoi{10.3847/1538-4357/aa7da1}

\bibitem[{{Andrews} {et~al.}(2018){Andrews}, {Huang}, {P{\'e}rez}, {Isella},
  {Dullemond}, {Kurtovic}, {Guzm{\'a}n}, {Carpenter}, {Wilner}, {Zhang}, {Zhu},
  {Birnstiel}, {Bai}, {Benisty}, {Hughes}, {{\"O}berg}, \& {Ricci}}]{DSHARPI}
{Andrews}, S.~M., {Huang}, J., {P{\'e}rez}, L.~M., {et~al.} 2018, \apjl, 869,
  L41, \dodoi{10.3847/2041-8213/aaf741}

\bibitem[{Armitage(2020)}]{armitage_2020}
Armitage, P.~J. 2020, Astrophysics of Planet Formation, 2nd edn. (Cambridge
  University Press), \dodoi{10.1017/9781108344227}

\bibitem[{{Astropy Collaboration} {et~al.}(2013){Astropy Collaboration},
  {Robitaille}, {Tollerud}, {Greenfield}, {Droettboom}, {Bray}, {Aldcroft},
  {Davis}, {Ginsburg}, {Price-Whelan}, {Kerzendorf}, {Conley}, {Crighton},
  {Barbary}, {Muna}, {Ferguson}, {Grollier}, {Parikh}, {Nair}, {Unther},
  {Deil}, {Woillez}, {Conseil}, {Kramer}, {Turner}, {Singer}, {Fox}, {Weaver},
  {Zabalza}, {Edwards}, {Azalee Bostroem}, {Burke}, {Casey}, {Crawford},
  {Dencheva}, {Ely}, {Jenness}, {Labrie}, {Lim}, {Pierfederici}, {Pontzen},
  {Ptak}, {Refsdal}, {Servillat}, \& {Streicher}}]{AstropyI}
{Astropy Collaboration}, {Robitaille}, T.~P., {Tollerud}, E.~J., {et~al.} 2013,
  \aap, 558, A33, \dodoi{10.1051/0004-6361/201322068}

\bibitem[{{Astropy Collaboration} {et~al.}(2018){Astropy Collaboration},
  {Price-Whelan}, {Sip{\H{o}}cz}, {G{\"u}nther}, {Lim}, {Crawford}, {Conseil},
  {Shupe}, {Craig}, {Dencheva}, {Ginsburg}, {VanderPlas}, {Bradley},
  {P{\'e}rez-Su{\'a}rez}, {de Val-Borro}, {Aldcroft}, {Cruz}, {Robitaille},
  {Tollerud}, {Ardelean}, {Babej}, {Bach}, {Bachetti}, {Bakanov}, {Bamford},
  {Barentsen}, {Barmby}, {Baumbach}, {Berry}, {Biscani}, {Boquien}, {Bostroem},
  {Bouma}, {Brammer}, {Bray}, {Breytenbach}, {Buddelmeijer}, {Burke},
  {Calderone}, {Cano Rodr{\'\i}guez}, {Cara}, {Cardoso}, {Cheedella}, {Copin},
  {Corrales}, {Crichton}, {D'Avella}, {Deil}, {Depagne}, {Dietrich}, {Donath},
  {Droettboom}, {Earl}, {Erben}, {Fabbro}, {Ferreira}, {Finethy}, {Fox},
  {Garrison}, {Gibbons}, {Goldstein}, {Gommers}, {Greco}, {Greenfield},
  {Groener}, {Grollier}, {Hagen}, {Hirst}, {Homeier}, {Horton}, {Hosseinzadeh},
  {Hu}, {Hunkeler}, {Ivezi{\'c}}, {Jain}, {Jenness}, {Kanarek}, {Kendrew},
  {Kern}, {Kerzendorf}, {Khvalko}, {King}, {Kirkby}, {Kulkarni}, {Kumar},
  {Lee}, {Lenz}, {Littlefair}, {Ma}, {Macleod}, {Mastropietro}, {McCully},
  {Montagnac}, {Morris}, {Mueller}, {Mumford}, {Muna}, {Murphy}, {Nelson},
  {Nguyen}, {Ninan}, {N{\"o}the}, {Ogaz}, {Oh}, {Parejko}, {Parley}, {Pascual},
  {Patil}, {Patil}, {Plunkett}, {Prochaska}, {Rastogi}, {Reddy Janga},
  {Sabater}, {Sakurikar}, {Seifert}, {Sherbert}, {Sherwood-Taylor}, {Shih},
  {Sick}, {Silbiger}, {Singanamalla}, {Singer}, {Sladen}, {Sooley},
  {Sornarajah}, {Streicher}, {Teuben}, {Thomas}, {Tremblay}, {Turner},
  {Terr{\'o}n}, {van Kerkwijk}, {de la Vega}, {Watkins}, {Weaver}, {Whitmore},
  {Woillez}, {Zabalza}, \& {Astropy Contributors}}]{AstropyII}
{Astropy Collaboration}, {Price-Whelan}, A.~M., {Sip{\H{o}}cz}, B.~M., {et~al.}
  2018, \aj, 156, 123, \dodoi{10.3847/1538-3881/aabc4f}

\bibitem[{{Bae} {et~al.}(2017){Bae}, {Zhu}, \&
  {Hartmann}}]{Bae..Zhu..Lee..2017}
{Bae}, J., {Zhu}, Z., \& {Hartmann}, L. 2017, \apj, 850, 201,
  \dodoi{10.3847/1538-4357/aa9705}

\bibitem[{{Bergin} {et~al.}(2014){Bergin}, {Cleeves}, {Crockett}, \&
  {Blake}}]{Bergin14}
{Bergin}, E.~A., {Cleeves}, L.~I., {Crockett}, N., \& {Blake}, G.~A. 2014,
  Faraday Discussions, 168, 61, \dodoi{10.1039/C4FD00003J}

\bibitem[{{Bergin} {et~al.}(2016){Bergin}, {Du}, {Cleeves}, {Blake}, {Schwarz},
  {Visser}, \& {Zhang}}]{Ted..et..al..2016}
{Bergin}, E.~A., {Du}, F., {Cleeves}, L.~I., {et~al.} 2016, \apj, 831, 101,
  \dodoi{10.3847/0004-637X/831/1/101}

\bibitem[{{Bergin} \& {Williams}(2018)}]{Ted..Williams..2018}
{Bergin}, E.~A., \& {Williams}, J.~P. 2018, arXiv e-prints, arXiv:1807.09631.
\newblock \doarXiv{1807.09631}

\bibitem[{{Bergner} {et~al.}(2019){Bergner}, {{\"O}berg}, {Bergin}, {Loomis},
  {Pegues}, \& {Qi}}]{Bergner..et..al..2019}
{Bergner}, J.~B., {{\"O}berg}, K.~I., {Bergin}, E.~A., {et~al.} 2019, \apj,
  876, 25, \dodoi{10.3847/1538-4357/ab141e}

\bibitem[{{Bergner} {et~al.}(2020){Bergner}, {{\"O}berg}, {Bergin}, {Andrews},
  {Blake}, {Carpenter}, {Cleeves}, {Guzm{\'a}n}, {Huang}, {J{\o}rgensen}, {Qi},
  {Schwarz}, {Williams}, \& {Wilner}}]{Bergner..et..al..2020}
---. 2020, \apj, 898, 97, \dodoi{10.3847/1538-4357/ab9e71}

\bibitem[{{Bethell} \& {Bergin}(2011)}]{Bethell..Bergin}
{Bethell}, T.~J., \& {Bergin}, E.~A. 2011, \apj, 740, 7,
  \dodoi{10.1088/0004-637X/740/1/7}

\bibitem[{{Birnstiel} {et~al.}(2010){Birnstiel}, {Dullemond}, \&
  {Brauer}}]{Birnstiel..et..al..2010}
{Birnstiel}, T., {Dullemond}, C.~P., \& {Brauer}, F. 2010, \aap, 513, A79,
  \dodoi{10.1051/0004-6361/200913731}

\bibitem[{{Birnstiel} {et~al.}(2012){Birnstiel}, {Klahr}, \&
  {Ercolano}}]{Birnstiel..et..al..2012}
{Birnstiel}, T., {Klahr}, H., \& {Ercolano}, B. 2012, \aap, 539, A148,
  \dodoi{10.1051/0004-6361/201118136}

\bibitem[{{Bosman} {et~al.}(2021{\natexlab{a}}){Bosman}, {Alarcon}, {Zhang}, \&
  {Bergin}}]{Bosman..2021}
{Bosman}, A.~D., {Alarcon}, F., {Zhang}, K., \& {Bergin}, E.~A.
  2021{\natexlab{a}}, arXiv e-prints, arXiv:2101.12502.
\newblock \doarXiv{2101.12502}

\bibitem[{{Bosman} {et~al.}(2021{\natexlab{b}}){Bosman}, {Bergin}, {Loomis},
  {Andrews}, {van 't Hoff}, {Teague}, {{\"O}berg}, {Guzm{\'a}n}, {Walsh},
  {Aikawa}, {Alarc{\'o}n}, {Bae}, {Bergner}, {Booth}, {Cataldi}, {Cleeves},
  {Czekala}, {Huang}, {Ilee}, {Law}, {Le Gal}, {Liu}, {Long}, {M{\'e}nard},
  {Nomura}, {P{\'e}rez}, {Qi}, \& {Schwarz}}]{bosman20_inner20au}
{Bosman}, A.~D., {Bergin}, E.~A., {Loomis}, R.~A., {et~al.} 2021{\natexlab{b}},
  arXiv e-prints, arXiv:2109.06223.
\newblock \doarXiv{2109.06223}

\bibitem[{{Bosman} {et~al.}(2021{\natexlab{c}}){Bosman}, {Alarc{\'o}n},
  {Bergin}, {Zhang}, {van 't Hoff}, {{\"O}berg}, {Guzm{\'a}n}, {Walsh},
  {Aikawa}, {Andrews}, {Bergner}, {Booth}, {Cataldi}, {Cleeves}, {Czekala},
  {Furuya}, {Huang}, {Ilee}, {Law}, {Le Gal}, {Liu}, {Long}, {Loomis},
  {M{\'e}nard}, {Nomura}, {Qi}, {Schwarz}, {Teague}, {Tsukagoshi}, {Yamato}, \&
  {Wilner}}]{bosman20_CtoO}
{Bosman}, A.~D., {Alarc{\'o}n}, F., {Bergin}, E.~A., {et~al.}
  2021{\natexlab{c}}, arXiv e-prints, arXiv:2109.06221.
\newblock \doarXiv{2109.06221}

\bibitem[{{Calahan} {et~al.}(2020){Calahan}, {Bergin}, {Zhang}, {Teague},
  {Cleeves}, {Bergner}, {Blake}, {Cazzoletti}, {Guzman}, {Hogerheijde},
  {Huang}, {Kama}, {Loomis}, {Oberg}, {van Dishoeck}, {Terwisscha van
  Scheltinga}, {Walsh}, {Wilner}, \& {Qi}}]{Calahan..et..al}
{Calahan}, J., {Bergin}, E., {Zhang}, K., {et~al.} 2020, arXiv e-prints,
  arXiv:2012.05927.
\newblock \doarXiv{2012.05927}

\bibitem[{{Cieza} {et~al.}(2019){Cieza}, {Ru{\'\i}z-Rodr{\'\i}guez}, {Hales},
  {Casassus}, {P{\'e}rez}, {Gonzalez-Ruilova}, {C{\'a}novas}, {Williams},
  {Zurlo}, {Ansdell}, {Avenhaus}, {Bayo}, {Bertrang}, {Christiaens}, {Dent},
  {Ferrero}, {Gamen}, {Olofsson}, {Orcajo}, {Pe{\~n}a Ram{\'\i}rez},
  {Principe}, {Schreiber}, \& {van der Plas}}]{Cieza..et..al..2019}
{Cieza}, L.~A., {Ru{\'\i}z-Rodr{\'\i}guez}, D., {Hales}, A., {et~al.} 2019,
  \mnras, 482, 698, \dodoi{10.1093/mnras/sty2653}

\bibitem[{{Cleeves}(2016)}]{Cleeves16}
{Cleeves}, L.~I. 2016, \apjl, 816, L21, \dodoi{10.3847/2041-8205/816/2/L21}

\bibitem[{{Cleeves} {et~al.}(2018){Cleeves}, {{\"O}berg}, {Wilner}, {Huang},
  {Loomis}, {Andrews}, \& {Guzman}}]{Cleeves..et..al..2018}
{Cleeves}, L.~I., {{\"O}berg}, K.~I., {Wilner}, D.~J., {et~al.} 2018, \apj,
  865, 155, \dodoi{10.3847/1538-4357/aade96}

\bibitem[{{Czekala} {et~al.}(2021){Czekala}, {Loomis}, {Teague}, {Booth},
  {Huang}, {Cataldi}, {Ilee}, {Law}, {Walsh}, {Bosman}, {Guzm{\'a}n}, {Le Gal},
  {{\"O}berg}, {Yamato}, {Aikawa}, {Andrews}, {Bae}, {Bergin}, {Bergner},
  {Cleeves}, {Kurtovic}, {M{\'e}nard}, {Nomura}, {P{\'e}rez}, {Qi}, {Schwarz},
  {Tsukagoshi}, {Waggoner}, {Wilner}, \& {Zhang}}]{czekala20}
{Czekala}, I., {Loomis}, R.~A., {Teague}, R., {et~al.} 2021, arXiv e-prints,
  arXiv:2109.06188.
\newblock \doarXiv{2109.06188}

\bibitem[{{Dong} {et~al.}(2017){Dong}, {Li}, {Chiang}, \&
  {Li}}]{Dong..et..al..2017}
{Dong}, R., {Li}, S., {Chiang}, E., \& {Li}, H. 2017, \apj, 843, 127,
  \dodoi{10.3847/1538-4357/aa72f2}

\bibitem[{{Dong} {et~al.}(2018){Dong}, {Li}, {Chiang}, \&
  {Li}}]{Dong..et..al..2018}
---. 2018, \apj, 866, 110, \dodoi{10.3847/1538-4357/aadadd}

\bibitem[{{Dong} {et~al.}(2015){Dong}, {Zhu}, \&
  {Whitney}}]{Dong..et..al..2015}
{Dong}, R., {Zhu}, Z., \& {Whitney}, B. 2015, \apj, 809, 93,
  \dodoi{10.1088/0004-637X/809/1/93}

\bibitem[{{Draine}(2003)}]{Draine..2003}
{Draine}, B.~T. 2003, \araa, 41, 241,
  \dodoi{10.1146/annurev.astro.41.011802.094840}

\bibitem[{{Du} \& {Bergin}(2014)}]{Fujun..Ted..RAC2D}
{Du}, F., \& {Bergin}, E.~A. 2014, \apj, 792, 2,
  \dodoi{10.1088/0004-637X/792/1/2}

\bibitem[{{Du} {et~al.}(2017){Du}, {Bergin}, {Hogerheijde}, {van Dishoeck},
  {Blake}, {Bruderer}, {Cleeves}, {Dominik}, {Fedele}, {Lis}, {Melnick},
  {Neufeld}, {Pearson}, \& {Y{\i}ld{\i}z}}]{Du17}
{Du}, F., {Bergin}, E.~A., {Hogerheijde}, M., {et~al.} 2017, \apj, 842, 98,
  \dodoi{10.3847/1538-4357/aa70ee}

\bibitem[{{Dullemond} {et~al.}(2012){Dullemond}, {Juhasz}, {Pohl}, {Sereshti},
  {Shetty}, {Peters}, {Commercon}, \& {Flock}}]{RADMC3D}
{Dullemond}, C.~P., {Juhasz}, A., {Pohl}, A., {et~al.} 2012, {RADMC-3D: A
  multi-purpose radiative transfer tool}.
\newblock \doeprint{1202.015}

\bibitem[{{Dullemond} {et~al.}(2018){Dullemond}, {Birnstiel}, {Huang},
  {Kurtovic}, {Andrews}, {Guzm{\'a}n}, {P{\'e}rez}, {Isella}, {Zhu}, {Benisty},
  {Wilner}, {Bai}, {Carpenter}, {Zhang}, \& {Ricci}}]{Dullemond..DSHARP}
{Dullemond}, C.~P., {Birnstiel}, T., {Huang}, J., {et~al.} 2018, \apjl, 869,
  L46, \dodoi{10.3847/2041-8213/aaf742}

\bibitem[{{Favre} {et~al.}(2019){Favre}, {Fedele}, {Maud}, {Booth}, {Tazzari},
  {Miotello}, {Testi}, {Semenov}, \& {Bruderer}}]{Favre..et..al..2019}
{Favre}, C., {Fedele}, D., {Maud}, L., {et~al.} 2019, \apj, 871, 107,
  \dodoi{10.3847/1538-4357/aaf80c}

\bibitem[{{Fedele} {et~al.}(2018){Fedele}, {Tazzari}, {Booth}, {Testi},
  {Clarke}, {Pascucci}, {Kospal}, {Semenov}, {Bruderer}, {Henning}, \&
  {Teague}}]{Fedele..et..al..2018}
{Fedele}, D., {Tazzari}, M., {Booth}, R., {et~al.} 2018, \aap, 610, A24,
  \dodoi{10.1051/0004-6361/201731978}

\bibitem[{{Fogel} {et~al.}(2011){Fogel}, {Bethell}, {Bergin}, {Calvet}, \&
  {Semenov}}]{Fogel..et..al..2011}
{Fogel}, J. K.~J., {Bethell}, T.~J., {Bergin}, E.~A., {Calvet}, N., \&
  {Semenov}, D. 2011, \apj, 726, 29, \dodoi{10.1088/0004-637X/726/1/29}

\bibitem[{{Gail} \& {Trieloff}(2017)}]{Gail17}
{Gail}, H.-P., \& {Trieloff}, M. 2017, \aap, 606, A16,
  \dodoi{10.1051/0004-6361/201730480}

\bibitem[{{Guzm{\'a}n} {et~al.}(2018){Guzm{\'a}n}, {Huang}, {Andrews},
  {Isella}, {P{\'e}rez}, {Carpenter}, {Dullemond}, {Ricci}, {Birnstiel},
  {Zhang}, {Zhu}, {Bai}, {Benisty}, {{\"O}berg}, \&
  {Wilner}}]{Vivi..et..al2018}
{Guzm{\'a}n}, V.~V., {Huang}, J., {Andrews}, S.~M., {et~al.} 2018, \apjl, 869,
  L48, \dodoi{10.3847/2041-8213/aaedae}

\bibitem[{{Guzm{\'a}n} {et~al.}(2021){Guzm{\'a}n}, {Bergner}, {Law}, {Oberg},
  {Walsh}, {Cataldi}, {Aikawa}, {Bergin}, {Czekala}, {Huang}, {Andrews},
  {Loomis}, {Zhang}, {Le Gal}, {Alarc{\'o}n}, {Ilee}, {Teague}, {Cleeves},
  {Wilner}, {Long}, {Schwarz}, {Bosman}, {P{\'e}rez}, {M{\'e}nard}, \&
  {Liu}}]{guzman20}
{Guzm{\'a}n}, V.~V., {Bergner}, J.~B., {Law}, C.~J., {et~al.} 2021, arXiv
  e-prints, arXiv:2109.06391.
\newblock \doarXiv{2109.06391}

\bibitem[{{Hasegawa} {et~al.}(1992){Hasegawa}, {Herbst}, \&
  {Leung}}]{Hasegawa..1992}
{Hasegawa}, T.~I., {Herbst}, E., \& {Leung}, C.~M. 1992, \apjs, 82, 167,
  \dodoi{10.1086/191713}

\bibitem[{{Henning} \& {Stognienko}(1996)}]{Henning..1996}
{Henning}, T., \& {Stognienko}, R. 1996, \aap, 311, 291

\bibitem[{{Henning} {et~al.}(2010){Henning}, {Semenov}, {Guilloteau}, {Dutrey},
  {Hersant}, {Wakelam}, {Chapillon}, {Launhardt}, {Pi{\'e}tu}, \&
  {Schreyer}}]{Henning..et..al..2010}
{Henning}, T., {Semenov}, D., {Guilloteau}, S., {et~al.} 2010, \apj, 714, 1511,
  \dodoi{10.1088/0004-637X/714/2/1511}

\bibitem[{{Hogerheijde} {et~al.}(2011){Hogerheijde}, {Bergin}, {Brinch},
  {Cleeves}, {Fogel}, {Blake}, {Dominik}, {Lis}, {Melnick}, {Neufeld},
  {Pani{\'c}}, {Pearson}, {Kristensen}, {Y{\i}ld{\i}z}, \& {van
  Dishoeck}}]{Hogerheijde11}
{Hogerheijde}, M.~R., {Bergin}, E.~A., {Brinch}, C., {et~al.} 2011, Science,
  334, 338, \dodoi{10.1126/science.1208931}

\bibitem[{{Huang} {et~al.}(2016){Huang}, {{\"O}berg}, \&
  {Andrews}}]{Jane..2016}
{Huang}, J., {{\"O}berg}, K.~I., \& {Andrews}, S.~M. 2016, \apjl, 823, L18,
  \dodoi{10.3847/2041-8205/823/1/L18}

\bibitem[{{Huang} {et~al.}(2018){Huang}, {Andrews}, {Dullemond}, {Isella},
  {P{\'e}rez}, {Guzm{\'a}n}, {{\"O}berg}, {Zhu}, {Zhang}, {Bai}, {Benisty},
  {Birnstiel}, {Carpenter}, {Hughes}, {Ricci}, {Weaver}, \&
  {Wilner}}]{DSHARPII}
{Huang}, J., {Andrews}, S.~M., {Dullemond}, C.~P., {et~al.} 2018, \apjl, 869,
  L42, \dodoi{10.3847/2041-8213/aaf740}

\bibitem[{{Kamp} {et~al.}(2011){Kamp}, {Woitke}, {Pinte}, {Tilling}, {Thi},
  {Menard}, {Duchene}, \& {Augereau}}]{Kamp..et..al..2011}
{Kamp}, I., {Woitke}, P., {Pinte}, C., {et~al.} 2011, \aap, 532, A85,
  \dodoi{10.1051/0004-6361/201016399}

\bibitem[{{Kanagawa} {et~al.}(2016){Kanagawa}, {Muto}, {Tanaka}, {Tanigawa},
  {Takeuchi}, {Tsukagoshi}, \& {Momose}}]{Kanagawa..et..al..2016}
{Kanagawa}, K.~D., {Muto}, T., {Tanaka}, H., {et~al.} 2016, \pasj, 68, 43,
  \dodoi{10.1093/pasj/psw037}

\bibitem[{{Kanagawa} {et~al.}(2017){Kanagawa}, {Tanaka}, {Muto}, \&
  {Tanigawa}}]{Kanagawa..et..al..2017}
{Kanagawa}, K.~D., {Tanaka}, H., {Muto}, T., \& {Tanigawa}, T. 2017, \pasj, 69,
  97, \dodoi{10.1093/pasj/psx114}

\bibitem[{{Kastner} {et~al.}(2014){Kastner}, {Hily-Blant}, {Rodriguez},
  {Punzi}, \& {Forveille}}]{Kastner14}
{Kastner}, J.~H., {Hily-Blant}, P., {Rodriguez}, D.~R., {Punzi}, K., \&
  {Forveille}, T. 2014, \apj, 793, 55, \dodoi{10.1088/0004-637X/793/1/55}

\bibitem[{{Kastner} {et~al.}(2015){Kastner}, {Qi}, {Gorti}, {Hily-Blant},
  {Oberg}, {Forveille}, {Andrews}, \& {Wilner}}]{Kastner15}
{Kastner}, J.~H., {Qi}, C., {Gorti}, U., {et~al.} 2015, \apj, 806, 75,
  \dodoi{10.1088/0004-637X/806/1/75}

\bibitem[{{Klarmann} {et~al.}(2018){Klarmann}, {Ormel}, \&
  {Dominik}}]{Klarmann18}
{Klarmann}, L., {Ormel}, C.~W., \& {Dominik}, C. 2018, \aap, 618, L1,
  \dodoi{10.1051/0004-6361/201833719}

\bibitem[{{Krijt} {et~al.}(2020){Krijt}, {Bosman}, {Zhang}, {Schwarz},
  {Ciesla}, \& {Bergin}}]{Krijt..et..al..2020}
{Krijt}, S., {Bosman}, A.~D., {Zhang}, K., {et~al.} 2020, \apj, 899, 134,
  \dodoi{10.3847/1538-4357/aba75d}

\bibitem[{{Law} {et~al.}(2021{\natexlab{a}}){Law}, {Loomis}, {Teague},
  {{\"O}berg}, {Czekala}, {Andrews}, {Huang}, {Aikawa}, {Alarc{\'o}n}, {Bae},
  {Bergin}, {Bergner}, {Boehler}, {Booth}, {Bosman}, {Calahan}, {Cataldi},
  {Cleeves}, {Furuya}, {Guzm{\'a}n}, {Ilee}, {Le Gal}, {Liu}, {Long},
  {M{\'e}nard}, {Nomura}, {Qi}, {Schwarz}, {Sierra}, {Tsukagoshi}, {Yamato},
  {van't Hoff}, {Walsh}, {Wilner}, \& {Zhang}}]{law20_rad}
{Law}, C.~J., {Loomis}, R.~A., {Teague}, R., {et~al.} 2021{\natexlab{a}}, arXiv
  e-prints, arXiv:2109.06210.
\newblock \doarXiv{2109.06210}

\bibitem[{{Law} {et~al.}(2021{\natexlab{b}}){Law}, {Teague}, {Loomis}, {Bae},
  {{\"O}berg}, {Czekala}, {Andrews}, {Aikawa}, {Alarc{\'o}n}, {Bergin},
  {Bergner}, {Booth}, {Bosman}, {Calahan}, {Cataldi}, {Cleeves}, {Furuya},
  {Guzm{\'a}n}, {Huang}, {Ilee}, {Le Gal}, {Liu}, {Long}, {M{\'e}nard},
  {Nomura}, {P{\'e}rez}, {Qi}, {Schwarz}, {Soto}, {Tsukagoshi}, {Yamato},
  {van't Hoff}, {Walsh}, {Wilner}, \& {Zhang}}]{law20_surf}
{Law}, C.~J., {Teague}, R., {Loomis}, R.~A., {et~al.} 2021{\natexlab{b}}, arXiv
  e-prints, arXiv:2109.06217.
\newblock \doarXiv{2109.06217}

\bibitem[{{Long} {et~al.}(2018){Long}, {Pinilla}, {Herczeg}, {Harsono},
  {Dipierro}, {Pascucci}, {Hendler}, {Tazzari}, {Ragusa}, {Salyk}, {Edwards},
  {Lodato}, {van de Plas}, {Johnstone}, {Liu}, {Boehler}, {Cabrit}, {Manara},
  {Menard}, {Mulders}, {Nisini}, {Fischer}, {Rigliaco}, {Banzatti}, {Avenhaus},
  \& {Gully-Santiago}}]{Long..et..al..2018}
{Long}, F., {Pinilla}, P., {Herczeg}, G.~J., {et~al.} 2018, \apj, 869, 17,
  \dodoi{10.3847/1538-4357/aae8e1}

\bibitem[{{Lynden-Bell} \& {Pringle}(1974)}]{Lynden-Bell_1974}
{Lynden-Bell}, D., \& {Pringle}, J.~E. 1974, \mnras, 168, 603,
  \dodoi{10.1093/mnras/168.3.603}

\bibitem[{{Mathis} {et~al.}(1977){Mathis}, {Rumpl}, \&
  {Nordsieck}}]{Mathis..et..al..1977}
{Mathis}, J.~S., {Rumpl}, W., \& {Nordsieck}, K.~H. 1977, \apj, 217, 425,
  \dodoi{10.1086/155591}

\bibitem[{{McElroy} {et~al.}(2013){McElroy}, {Walsh}, {Markwick}, {Cordiner},
  {Smith}, \& {Millar}}]{McElroy..et..al..2012}
{McElroy}, D., {Walsh}, C., {Markwick}, A.~J., {et~al.} 2013, \aap, 550, A36,
  \dodoi{10.1051/0004-6361/201220465}

\bibitem[{{Miotello} {et~al.}(2014){Miotello}, {Bruderer}, \& {van
  Dishoeck}}]{Miotello..et..al..2014}
{Miotello}, A., {Bruderer}, S., \& {van Dishoeck}, E.~F. 2014, \aap, 572, A96,
  \dodoi{10.1051/0004-6361/201424712}

\bibitem[{{Miotello} {et~al.}(2019){Miotello}, {Facchini}, {van Dishoeck},
  {Cazzoletti}, {Testi}, {Williams}, {Ansdell}, {van Terwisga}, \& {van der
  Marel}}]{Miotello..et..al..2019}
{Miotello}, A., {Facchini}, S., {van Dishoeck}, E.~F., {et~al.} 2019, \aap,
  631, A69, \dodoi{10.1051/0004-6361/201935441}

\bibitem[{{Molyarova} {et~al.}(2017){Molyarova}, {Akimkin}, {Semenov},
  {Henning}, {Vasyunin}, \& {Wiebe}}]{Molyarova..et..al..2017}
{Molyarova}, T., {Akimkin}, V., {Semenov}, D., {et~al.} 2017, \apj, 849, 130,
  \dodoi{10.3847/1538-4357/aa9227}

\bibitem[{{Nieva} \& {Przybilla}(2012)}]{Nieva..Przybilla..2012}
{Nieva}, M.~F., \& {Przybilla}, N. 2012, \aap, 539, A143,
  \dodoi{10.1051/0004-6361/201118158}

\bibitem[{{{\"O}berg} {et~al.}(2009{\natexlab{a}}){{\"O}berg}, {Linnartz},
  {Visser}, \& {van Dishoeck}}]{Oberg..et.al..2009}
{{\"O}berg}, K.~I., {Linnartz}, H., {Visser}, R., \& {van Dishoeck}, E.~F.
  2009{\natexlab{a}}, \apj, 693, 1209, \dodoi{10.1088/0004-637X/693/2/1209}

\bibitem[{{{\"O}berg} {et~al.}(2009{\natexlab{b}}){{\"O}berg}, {van Dishoeck},
  \& {Linnartz}}]{Oberg..2009b}
{{\"O}berg}, K.~I., {van Dishoeck}, E.~F., \& {Linnartz}, H.
  2009{\natexlab{b}}, \aap, 496, 281, \dodoi{10.1051/0004-6361/200810207}

\bibitem[{{Oberg} {et~al.}(2021){Oberg}, {Guzman}, {Walsh}, {Aikawa}, {Bergin},
  {Law}, {Loomis}, {Alarcon}, {Andrews}, {Bae}, {Bergner}, {Boehler}, {Booth},
  {Bosman}, {Calahan}, {Cataldi}, {Cleeves}, {Czekala}, {Furuya}, {Huang},
  {Ilee}, {Kurtovic}, {Le Gal}, {Liu}, {Long}, {Menard}, {Nomura}, {Perez},
  {Qi}, {Schwarz}, {Sierra}, {Teague}, {Tsukagoshi}, {Yamato}, {van 't Hoff},
  {Waggoner}, {Wilner}, \& {Zhang}}]{oberg20}
{Oberg}, K.~I., {Guzman}, V.~V., {Walsh}, C., {et~al.} 2021, arXiv e-prints,
  arXiv:2109.06268.
\newblock \doarXiv{2109.06268}

\bibitem[{{Pinilla} {et~al.}(2012){Pinilla}, {Benisty}, \&
  {Birnstiel}}]{Pinilla..et..al..2012}
{Pinilla}, P., {Benisty}, M., \& {Birnstiel}, T. 2012, \aap, 545, A81,
  \dodoi{10.1051/0004-6361/201219315}

\bibitem[{{Rab} {et~al.}(2020){Rab}, {Kamp}, {Dominik}, {Ginski}, {Muro-Arena},
  {Thi}, {Waters}, \& {Woitke}}]{Rab..2020}
{Rab}, C., {Kamp}, I., {Dominik}, C., {et~al.} 2020, \aap, 642, A165,
  \dodoi{10.1051/0004-6361/202038712}

\bibitem[{{Reboussin} {et~al.}(2015){Reboussin}, {Wakelam}, {Guilloteau},
  {Hersant}, \& {Dutrey}}]{Reboussin15}
{Reboussin}, L., {Wakelam}, V., {Guilloteau}, S., {Hersant}, F., \& {Dutrey},
  A. 2015, \aap, 579, A82, \dodoi{10.1051/0004-6361/201525885}

\bibitem[{{Rosotti} {et~al.}(2020){Rosotti}, {Teague}, {Dullemond}, {Booth}, \&
  {Clarke}}]{Rosotti..et..al}
{Rosotti}, G.~P., {Teague}, R., {Dullemond}, C., {Booth}, R.~A., \& {Clarke},
  C.~J. 2020, \mnras, 495, 173, \dodoi{10.1093/mnras/staa1170}

\bibitem[{{Schwarz} {et~al.}(2019){Schwarz}, {Bergin}, {Cleeves}, {Zhang},
  {{\"O}berg}, {Blake}, \& {Anderson}}]{Kamber..et..al..2019}
{Schwarz}, K.~R., {Bergin}, E.~A., {Cleeves}, L.~I., {et~al.} 2019, \apj, 877,
  131, \dodoi{10.3847/1538-4357/ab1c5e}

\bibitem[{{Smirnov-Pinchukov} {et~al.}(2020){Smirnov-Pinchukov}, {Semenov},
  {Akimkin}, \& {Henning}}]{Smirnov..et..al..2020}
{Smirnov-Pinchukov}, G.~V., {Semenov}, D.~A., {Akimkin}, V.~V., \& {Henning},
  T. 2020, arXiv e-prints, arXiv:2009.09962.
\newblock \doarXiv{2009.09962}

\bibitem[{{Teague} {et~al.}(2018){Teague}, {Bae}, {Birnstiel}, \&
  {Bergin}}]{Teague..et..al..2018}
{Teague}, R., {Bae}, J., {Birnstiel}, T., \& {Bergin}, E.~A. 2018, \apj, 868,
  113, \dodoi{10.3847/1538-4357/aae836}

\bibitem[{{van der Marel} {et~al.}(2018){van der Marel}, {Williams}, \&
  {Bruderer}}]{Nienke..et..al..2018}
{van der Marel}, N., {Williams}, J.~P., \& {Bruderer}, S. 2018, \apjl, 867,
  L14, \dodoi{10.3847/2041-8213/aae88e}

\bibitem[{{van Dishoeck} {et~al.}(2021){van Dishoeck}, {Kristensen}, {Mottram},
  {Benz}, {Bergin}, {Caselli}, {Herpin}, {Hogerheijde}, {Johnstone}, {Liseau},
  {Nisini}, {Tafalla}, {van der Tak}, {Wyrowski}, {Baudry}, {Benedettini},
  {Bjerkeli}, {Blake}, {Braine}, {Bruderer}, {Cabrit}, {Cernicharo}, {Choi},
  {Coutens}, {de Graauw}, {Dominik}, {Fedele}, {Fich}, {Fuente}, {Furuya},
  {Goicoechea}, {Harsono}, {Helmich}, {Herczeg}, {Jacq}, {Karska}, {Kaufman},
  {Keto}, {Lamberts}, {Larsson}, {Leurini}, {Lis}, {Melnick}, {Neufeld},
  {Pagani}, {Persson}, {Shipman}, {Taquet}, {van Kempen}, {Walsh}, {Wampfler},
  {Y{\i}ld{\i}z}, \& {WISH Team}}]{Ewine..2021}
{van Dishoeck}, E.~F., {Kristensen}, L.~E., {Mottram}, J.~C., {et~al.} 2021,
  \aap, 648, A24, \dodoi{10.1051/0004-6361/202039084}

\bibitem[{Warren \& Brandt(2008)}]{Warren..and..Brandt}
Warren, S.~G., \& Brandt, R.~E. 2008, Journal of Geophysical Research:
  Atmospheres, 113, \dodoi{10.1029/2007JD009744}

\bibitem[{{Williams} \& {Best}(2014)}]{Williams..Best..2015}
{Williams}, J.~P., \& {Best}, W. M.~J. 2014, \apj, 788, 59,
  \dodoi{10.1088/0004-637X/788/1/59}

\bibitem[{{Wilson}(1999)}]{Wilson..et..al..1999}
{Wilson}, T.~L. 1999, Reports on Progress in Physics, 62, 143,
  \dodoi{10.1088/0034-4885/62/2/002}

\bibitem[{{Wilson} \& {Rood}(1994)}]{Wilson..and..Rood}
{Wilson}, T.~L., \& {Rood}, R. 1994, \araa, 32, 191,
  \dodoi{10.1146/annurev.aa.32.090194.001203}

\bibitem[{{Yu} {et~al.}(2017){Yu}, {Evans}, {Dodson-Robinson}, {Willacy}, \&
  {Turner}}]{Yu..et..al..2017}
{Yu}, M., {Evans}, Neal~J., I., {Dodson-Robinson}, S.~E., {Willacy}, K., \&
  {Turner}, N.~J. 2017, \apj, 841, 39, \dodoi{10.3847/1538-4357/aa6e4c}

\bibitem[{{Zhang} {et~al.}(2021){Zhang}, {Booth}, {Law}, {Bosman}, {Schwarz},
  {Bergin}, {{\"O}berg}, {Andrews}, {Guzm{\'a}n}, {Walsh}, {Qi}, {van 't Hoff},
  {Long}, {Wilner}, {Huang}, {Czekala}, {Ilee}, {Cataldi}, {Bergner}, {Aikawa},
  {Teague}, {Bae}, {Loomis}, {Calahan}, {Alarc{\'o}n}, {M{\'e}nard}, {Le Gal},
  {Sierra}, {Yamato}, {Nomura}, {Tsukagoshi}, {P{\'e}rez}, {Trapman}, {Liu}, \&
  {Furuya}}]{zhang20_maps}
{Zhang}, K., {Booth}, A.~S., {Law}, C.~J., {et~al.} 2021, arXiv e-prints,
  arXiv:2109.06233.
\newblock \doarXiv{2109.06233}

\bibitem[{{Zhang} {et~al.}(2018){Zhang}, {Zhu}, {Huang}, {Guzm{\'a}n},
  {Andrews}, {Birnstiel}, {Dullemond}, {Carpenter}, {Isella}, {P{\'e}rez},
  {Benisty}, {Wilner}, {Baruteau}, {Bai}, \& {Ricci}}]{Zhang..DSHARP}
{Zhang}, S., {Zhu}, Z., {Huang}, J., {et~al.} 2018, \apjl, 869, L47,
  \dodoi{10.3847/2041-8213/aaf744}

\end{thebibliography}
\bibliographystyle{aasjournal}

\end{document}